\documentclass[12pt]{article}
\pdfoutput=1
\usepackage{jheppub}
\usepackage{amsmath,amsthm,amssymb}
\usepackage{graphicx}
\usepackage[usenames,dvipsnames]{xcolor}
\usepackage{calligra}
\DeclareMathAlphabet{\mathcalligra}{T1}{calligra}{m}{n}
\DeclareFontShape{T1}{calligra}{m}{n}{<->s*[2.2]callig15}{}
\newcommand{\scripty}[1]{\ensuremath{\mathcalligra{#1}}}

\numberwithin{equation}{section}

\makeatletter
\def\@fpheader{\relax}
\makeatother

\def\p{\partial}
\def\be{\begin{equation}}
\def\ee{\end{equation}}
\def\ba{\begin{array}}
\def\ea{\end{array}}
\def\beq{\begin{eqnarray}}
\def\eeq{\end{eqnarray}}
\def\bea{\begin{eqnarray}}
\def\eea{\end{eqnarray}}
\newcommand\eref[1]{\eqref{#1}}
\newcommand\nn{\nonumber}

\newcommand\comment[1]{}

\title{Generalized Hot Attractors}

\author[*]{Kevin Goldstein,}
\author[*\dagger]{Vishnu Jejjala,}
\author[*]{James Junior Mashiyane,}
\author[\ddagger]{Suresh Nampuri}

\affiliation[*]{Mandelstam Institute for Theoretical Physics, School of Physics, NITheP, and CoE-MaSS, University of the Witwatersrand, 1 Jan Smuts Avenue, Johannesburg, WITS 2050, South Africa}
\affiliation[\dagger]{David Rittenhouse Laboratory, 209 S.\ 33rd Street, University of Pennsylvania,
Philadelphia, PA 19104, USA}
\affiliation[\ddagger]{Center for Mathematical Analysis, Geometry, and Dynamical Systems, Instituto Superior T\'ecnico, Universidade de Lisboa, Av.\ Rovisco Pais, 1049-001 Lisboa, Portugal}

\emailAdd{kevin.goldstein@wits.ac.za}
\emailAdd{vishnu@neo.phys.wits.ac.za}
\emailAdd{James.Mashiyane@students.wits.ac.za}
\emailAdd{snampuri@math.tecnico.ulisboa.pt}

\abstract{
Non-extremal black holes are endowed with geometric invariants related to their horizon areas.
We extend earlier work on hot attractor black holes to higher dimensions and add a scalar potential.
In addition to the event and Cauchy horizons, when we complexify the radial coordinate, non-extremal black holes will generically have other horizons as well.
We prove that the product of all of the horizon areas is independent of variations of the asymptotic moduli further generalizing the attractor mechanism for extremal black holes.
In the presence of a scalar potential, as typically appears in gauged supergravity, we find that the product of horizon areas is not necessarily the geometric mean of the extremal area, however.
We outline the derivation of horizon invariants for stationary backgrounds.
}

\begin{document}

\maketitle

\section{Introduction}

String theory is evidently the correct language for studying black holes.
As the strength of gravity is increased by tuning the Newton coupling, quantities such as the sizes of orbits within the solar system or the radius of neutron stars become smaller.
The Schwarzschild radius, however, scales with the Newton coupling --- $r_h = 2G_N M$ --- so event horizons grow.
Black holes and black hole microstates in string theory exhibit this property~\cite{Horowitz:1996nw,BSV}.

Black holes obey the laws of thermodynamics~\cite{Bekenstein:1973ur,Bardeen:1973gs,Bekenstein:1974ax,Hawking:1974sw}.
The identification of the area of the event horizon with the entropy requires us to distinguish $e^{S_\mathrm{BH}}$ microstates with the same charges as the black hole.
String theory accomplishes this task for a few systems, in particular for $\frac14$-BPS configurations of D$1$-branes and D$5$-branes.
Due to maximum supersymmetry, the counting at weak coupling~\cite{sv1} equals the counting at strong coupling~\cite{sv2}.
Something similar happens for the $\frac12$-BPS superstar in AdS$_5\times S^5$ where microstates are sourced by giant gravitons (D$3$-branes wrapping spheres)~\cite{Lin:2004nb,Balasubramanian:2005mg}.
The construction of explicit states relies on the gauge/gravity correspondence~\cite{adscft1,adscft2,adscft3}.
The matching works exactly for certain small black holes~\cite{Dabholkar:2004yr}, but state counting when the area of the event horizon is finite in supergravity is considerably more difficult~\cite{Sen:2009bm,Mandal:2010cj}.

There is evidence, however, that the intuition gained from studying supersymmetric and extremal systems extends to non-supersymmetric and non-extremal settings.
Though insufficient in number to account for the entropy, non-extremal states in supergravity with identical charges to a black hole have been constructed in the D$1$-D$5$-P system and various generalizations thereof~\cite{jmart,Giusto:2007tt,bgrw,ccdm,dgr,bnw,nie,bcvv,ctv,bbkt}.
Similarly, analysis of near-extremal black holes in AdS$_5$ has led to consideration of a gas of defects in BPS operators with conformal dimensions of $O(N^2)$ as a dual description of their spacetime geometries~\cite{Balasubramanian:2005kk,Balasubramanian:2007bs}.
The Kerr/CFT correspondence moreover suggests that a two dimensional conformal field theory counts the entropy of near extremal astrophysical black holes~\cite{kerrcft,Castro:2009jf,Bredberg:2011hp,Compere:2012jk}.
The Cardy formula~\cite{cardy1,cardy2,cardy3}, which is the leading term in the high temperature expansion of the partition function, applies well beyond its na\"{\i}ve regime of validity~\cite{ntt,jn,Belin:2016yll}.

The generalized attractor mechanism in $\mathcal{N}=2$ supergravity embeds in string theory and links extremal and non-extremal black holes.
The attractor mechanism, as originally developed~\cite{attr1,attr2,attr3,attr4,attr5,hep-th/0506177,hep-th/0507096,hep-th/0512138,hep-th/0606244,hep-th/0611143,0708.1270}, operates on extremal solutions and explains that the horizon area and therefore the entropy are independent of background moduli.
The presence of an AdS$_2$ throat in the near horizon region damps away asymptotic fluctuations in the scalars and fixes the moduli to attractor values at the horizon.
This has to happen because while moduli at infinity may be varied continuously, the entropy is a function of quantized charges, is computed using an index, and jumps discretely.
Decoupling of the inner throat from the asymptotic region ensures that on-shell values of the scalar fields at the horizon are fixed by the black hole's charges.
The area is calculated from evaluating the minimum of an effective potential for the scalars that measures the energy density in the electromagnetic fields.
Sen's entropy function~\cite{hep-th/0506177} computes the on-shell value of the Legendre transform of the action.
Crucially, the attractor mechanism depends solely on extremality and not on supersymmetry~\cite{hep-th/0507096}.

Non-extremal black holes can have an event horizon and a Cauchy horizon.
In $\mathcal{N}=2$ supergravity, the values of the scalar fields at the two horizons depend on the moduli at infinity.
Surprisingly, the attractor mechanism operates to fix robust invariants~\cite{gjn1} in this case as well.\footnote{
See~\cite{Charles:2015eha,Charles:2017dbr,Castro:2018hsc} for recent developments on universal (only model dependent) coefficients of the logarithmic correction to black hole entropy in specific theories.
See also~\cite{Hajian:2016iyp,Larsen:2018iou,Moitra:2018jqs}.}
The geometric mean of the areas of the inner and the outer horizons is the area of the event horizon of the extremal black hole obtained from taking a smooth zero temperature limit.
The observation that
\be
\label{geom}
A_+\, A_- = A_\mathrm{ext}^2
\ee
was initially phenomenological~\cite{OK,finn,cgp,cretc1}, and the fact that the right hand side of \eqref{geom} only depends on quantized charges suggests an underlying microscopic explanation of the moduli independence of the horizon products~\cite{cgp}.
The proof of \eqref{geom} in $\mathcal{N}=2$ supergravity  establishes the \textit{hot attractor mechanism}  and consequently generalizes the extremal attractor mechanism to non-extremal black holes.
Significantly, the proof does not invoke supersymmetry. 
Certain quantities that vanish by virtue of the attractor equations on extremal solutions now vanish when integrated over the so called Region~2 between the inner and the outer horizons~\cite{gjn1}.
This suggests that Region~2, which is decoupled from infinity, may explicate the entropy for these non-extremal black holes~\cite{gjn2}.\footnote{
See~\cite{Martinec} for a similar proposal in the context of the fuzzball program.}
Indeed, by fixing the scalar fields to their attractor values, non-extremal Reissner--Nordstr\"om black holes can be uplifted to BTZ black holes in AdS$_3\times S^2$ by means of a Harrison transformation~\cite{cl1,cl2,cg}.
In the limit where the temperature goes to zero, a mass gap develops and the familiar attractor mechanism for extremal solutions can be understood in terms of hot attractors~\cite{gjn2}.
In this limit, the $SL(2,\mathbb{R})$ symmetry in the moduli space of scalar fields near their attractor values becomes the AdS$_2$ symmetry in spacetime.

In the present work, we extend the hot attractor mechanism to more general settings. We look at the mechanism in higher dimensions and study in detail the effect of an additional scalar potential. Interestingly, we find that in the presence of a scalar potential, while the product of horizon areas is an invariant, it is not the geometric mean of the extremal area as in \eqref{geom}.\footnote{
This fact was found though not explicitly pointed out in~\cite{Toldo:2012ec}.
Various horizon invariants of  (A)dS Reissner--Nordstr\"om black holes were noted in~\cite{Wang:2013smb,Xu:2013zpa,Xu:2014qaa,Du:2014kpa}. }

The organization of the paper is as follows.
Section~\ref{sec:rev} reviews the attractor equations.
Section~\ref{sec:gd} shows how a change of coordinates allows us to construct hot attractors in higher dimensions.
Section~\ref{sec:infd} briefly explores the $d\to\infty$ limit of hot attractors.
In the presence of a scalar potential, we can have horizons whose radii are distributed at various points on the complex plane.
Section~\ref{sec:cplx} discusses invariants that can be constructed from products of horizon areas.
Section~\ref{sec:rotation} examines the case of stationary black holes and sketches the derivation of an invariant associated to the Killing horizons.
Finally, in Section~\ref{sec:disc} we summarize and discuss our results.

\section{Review of attractors: cold and hot}\label{sec:rev}
We begin by considering four dimensional gravity coupled to $U(1)$ gauge fields with the action 
\be
\label{act1}
S = \frac{1}{16\pi G}\int d^4x\ \sqrt{-G}\ \big( R - 2 g_{ij}(\phi) \partial_\mu\phi^i \partial^\mu\phi^j - f_{ab}(\phi) F_{\mu\nu}^a F^{b\ \mu\nu} - \frac12 \widetilde{f}_{ab}(\phi) F_{\mu\nu}^a F_{\rho\sigma}^b \epsilon^{\mu\nu\rho\sigma} -2 V_g(\phi)\big) ~,
\ee
where $G_{\mu\nu}$ is the spacetime metric, $g_{ij}$ is the metric in field space, and $V_g(\phi)$ is a scalar potential.  
With specific choices for $g_{ij}$, $f_{ab}$, $\tilde{f}_{ab}$ and $V_g$, \eqref{act1} corresponds to the bosonic part of an $\mathcal{N}=2$ gauged supergravity.
Taking the ansatz~\cite{hep-th/0507096}
\bea
\mathrm{d}s^2 &=& -a(r)^2\, dt^2 + a(r)^{-2}\, dr^2 + b(r)^2\, \mathrm{d}\Omega_2^2 ~, \\
F^a &=& f^{ab}(\phi) (Q_{eb}-\widetilde{f}_{bc}(\phi) Q^c_m) \frac{1}{b^2}\, dt\wedge dr + Q^a_m \sin\theta\, d\theta \wedge d\phi ~, \\
\phi^i &=& \phi^i(r) ~,
\eea
the equations of motion lead to
\bea
(a^2b^2)'' &=& 2\,-4\,V_g b^2\, ~, \label{eq:eom1} \\
\frac{b''}{b} &=& -\sum_i \big(\phi^{i\, \prime}\big)^2 ~, \label{eq:eom2} \\
\left(a^2b^2 \phi^{i\, \prime}\right)' &=&  \tfrac{1}{2}\left(\frac{\partial_i V_\mathrm{eff}}{ b^2} +\partial_i V_g b^2\right)~, \label{eq:eom3}
\eea
where primes denote differentiation with respect to the radial coordinate $r$, $\partial_i = \frac{\partial}{\partial \phi^i}$, and the energy density in the electromagnetic fields is characterized by an effective potential
\be
V_\mathrm{eff}(\phi) = f^{ab}(\phi)(Q_{ea}-\widetilde{f}_{ac}(\phi) Q^c_m)(Q_{eb}-\widetilde{f}_{bd}(\phi) Q^d_m) + f_{ab}(\phi) Q^a_m Q^b_m ~. \label{eq:veff}
\ee
Here, $Q_{ea}$ and $Q^a_m$ are electric and magnetic charges carried by gauge fields and $f^{ab}(\phi)$ is the matrix inverse of $f_{ab}(\phi)$.
The non-linear coupled second order differential equations involving the warp factors $a(r)$ and $b(r)$ in the metric and the scalars $\phi^i(r)$ are supplemented by a first order energy constraint
\be
-1 + a^2b^{\prime2}+\frac12 (a^2)'(b^2)' = -\frac{V_\mathrm{eff}(\phi)}{b^2} - b^2 V_g + a^2b^2 \sum_i \big(\phi^{i\,\prime}\big)^2 ~. \label{eq:hc}
\ee

The spherically symmetric solutions to the equations of motion that satisfy the constraint~\eref{eq:hc} can be either extremal or non-extremal.
The extremality condition allows us to write the mass in terms of the charges.
We emphasize that in this context the terms \textit{supersymmetric} and \textit{extremal} are not synonyms.
Supersymmetric solutions are of course extremal, but non-supersymmetric extremal solutions also exist and our discussion applies to them as well.

When $V_g=0$, \eqref{eq:eom1} can be immediately integrated to give $a^2 b^2=(r-r_-)(r-r_+)$.
As usual, horizons are at roots $r_\pm$ where $g^{rr} = a^2(r) = 0$.
The warp factor $b(r_\pm)$ tells us the radius of the $S^2$ at the horizons.
The roots $r_\pm$ coincide for extremal solutions.
The singularity corresponds to the roots of $b^2(r) = 0$.
For extremal solutions, because the presence of a double horizon leads to cancelation of terms in~\eref{eq:hc}, the Hamiltonian constraint simplifies to
\be
b^2_\mathrm{ext} = V_\mathrm{eff,min} ~, \label{eq:hc2}
\ee
where we use the subscript ``ext'' to denote quantities evaluated at a double horizon and the subscript `` min'' to denote quantities evaluated at a minimum of a potential.
In this case, since the near horizon geometry is AdS$_2\times S^2$, the attractor mechanism ensures that the moduli at infinity assume fixed values $\phi^i_0$ at the horizon, which, consistent with the ``no hair'' theorem, are determined by the charges~\cite{hep-th/0506177}.
From~\eref{eq:hc2}, the zero temperature solutions have an event horizon whose area is determined by the effective potential evaluated on the attractor values of the moduli.

For non-extremal black holes, the scalars at the inner and the outer horizons vary with the asymptotic values of the moduli.
Nevertheless, in~\cite{gjn1,gjn2} we show that from the existence of a smooth zero temperature limit that connects non-extremal black holes to extremal solutions, there are invariants.
In particular, the geometric mean of the areas of the inner and outer horizons of the finite temperature black hole is the area of the event horizon of the zero temperature black hole.

Also, for non-extremal solutions
\bea
\int_{r_-}^{r_+} dr\ \frac{\partial_i V_\mathrm{eff}(\phi)}{b^2} &=& 0 ~, \label{eq:int1} \\
\int_{r_-}^{r_+} dr\ \left( \frac{V_\mathrm{eff}(\phi)}{b^2} - 1 \right) &=& 0 ~. \label{eq:int2}
\eea
For extremal solutions, the integrands of~\eref{eq:int1} and~\eref{eq:int2} are zero since the moduli are fixed at the horizon to the attractor values at the minimum of the potential, whereas in the non-extremal case, the integrals of the same quantities over the inter-horizon region (Region~2) vanish.
Averaging over the region between the horizons with a uniform measure recapitulates the attractor equations for the extremal solution.
In the next section, we will derive similar expressions in arbitrary spacetime dimensions.

When $V_g\neq 0$, while \eqref{eq:eom1} becomes harder to integrate, solutions with constant scalars can be found \cite{Toldo:2012ec}.
For constant scalars,~\eref{eq:eom2} tells us that we can choose coordinates so that $b(r) = r$.
In \eqref{eq:eom3}, the terms on the right hand side then have different powers of $r$.
As the left hand side is zero, these two terms must vanish separately in order that the relation holds at any value of the radial coordinate.
We require that
\be
\label{min}
\partial_i V_\text{eff}= 0 ~, \qquad \p_i V_g =0 ~.
\ee
If there are $K$ scalar fields, these are $2K$ equations in $K$ variables.
Often, the scalars that contribute to $V_g$ and $V_\mathrm{eff}$ originate in different multiplets in supergravity.
There are cases in which the system is not over determined and both sets of constraints can be solved simultaneously.
As discussed in \cite{hep-th/0507096}, for black hole solutions, the extrema in \eqref{min} are required to be minima.     
Using this and substituting the solution for $b(r)$ into \eqref{eq:eom1} gives
\begin{equation}
\label{a2b2:quartic}
a^2 b^2 = \frac{(-V_{g,\text{min}})}{3}r^4+r^2-c_1 r+c_2 =\frac{(-V_{g,\text{min}})}{3}(r-r_+)(r-r_-)(r-\lambda)(r-\lambda^*)~,
\end{equation}
where $c_{1,2}$ are constants of integration (related to the mass and charges of the solution respectively) with 
\begin{eqnarray}
\label{const}
c_1 &=& \tfrac{1}{3}(r_++r_-)\left(3+(-V_{g,\text{min}})(r_+^2+r_-^2)\right) = 2 G M ~, \cr
c_2 &=& \tfrac{1}{3} r_+r_- \left(3+(-V_{g,\text{min}})(r_+^2+r_+r_-+r_-^2) \right) = V_\mathrm{eff,min}  ~,
\end{eqnarray}
and
\begin{equation}
\label{eq:lambda}
\lambda = -\frac{r_++r_-}{2}+ \tfrac{i}{2} \sqrt{\Xi}~, \qquad \text{where} ~~ \Xi=(r_++r_-)^2+2(r_+^2+r_-^2)+\tfrac{12}{(-V_{g,\text{min}})	}~.
\end{equation}

From \eqref{a2b2:quartic}, in requiring $g_{tt}$ to be timelike at infinity, we need $V_{g,\text{min}}$ to be negative, which means that the constant scalar solutions are essentially AdS Reissner--Nordstr\"om black holes. 
The negativity of $V_{g,\text{min}}$, also tells us, using \eqref{eq:lambda}, that $\lambda$ is not real (for real $r_\pm$). 

In general, when the scalars are not constant and $V_g\neq 0$, if we have an extremal solution with both real horizons coinciding, which is to say $a=a'=0$ at the horizon, then evaluating \eqref{eq:eom3} at the horizon gives
\begin{equation}
\label{gen_atr}
\frac{\partial_i V_\mathrm{eff}(\phi)}{b_\text{ext}^2}+b_\text{ext}^2\p_i V_g =0 ~,
\end{equation}
which can be solved by \eqref{min} or, in principle, by
\begin{equation}
\label{bsol2}
b^2_\text{ext}=\sqrt{-\frac{\p_i V_\text{eff}}{\p_i V_g}}
\end{equation}
for any $i$ for which \eqref{min} does not hold.
There is no obvious obstruction to solving \eqref{gen_atr} with scalars that satisfy \eqref{bsol2}, but we are not aware of any solutions to gauged supergravity that solve the attractor equations through this relation.
For the purposes of our analysis and to make contact with known solutions in the literature, we will restrict our attention to solutions which satisfy \eqref{min} at the horizon.
With this in mind, evaluating the Hamiltonian constraint \eqref{eq:hc} at the horizon gives
\begin{equation}
\label{gen_atr2}
\frac{V_\mathrm{eff,min}}{b^2_\text{ext}} + b_\text{ext}^2V_\text{$g$,min}= 1 ~.
\end{equation}
This tells us that\footnote{
Since $b_\text{ext}^2$ is positive, we only take the positive root of \eqref{gen_atr2}. }
\begin{equation}
\label{bext_gen}
b_\text{ext}^2=\frac{\sqrt{1+4 V_\mathrm{eff,min}(-V_\text{$g$,min})}-1}{2(-V_\text{$g$,min})}~.
\end{equation}
When we do not have an extremal solution,
by integrating between the horizons, we can generalize \eqref{eq:int1} and \eqref{eq:int2} to read
\bea
\int_{r_-}^{r_+} dr\ \left(\frac{\partial_i V_\mathrm{eff}}{b^2}+b^2\p_i V_g\right) &=& 0 ~, \label{eq:int3} \\
\int_{r_-}^{r_+} dr\ \left( \frac{V_\mathrm{eff}}{b^2} + b^2V_g- 1 \right) &=& 0 ~, \label{eq:int4}
\eea
which once again averages the attractor equations \eqref{gen_atr} and \eqref{gen_atr2} over Region~2.

\section{Attractors in general dimensions}\label{sec:gd}
In general dimensions,  we consider an action of the form
\be
S = \frac{1}{{16\pi G_d}} \int d^dx\ \sqrt{-G} \left(R - 2 g_{ij}(\phi) \partial\phi^i \partial\phi^j - f_{ab}(\phi)F^a F^b - 2V_g(\phi) \right) ~.
\label{eq:ACTION}
\ee
In order to have solutions that enjoy an $SO(d-2)$ spherical symmetry, we restrict to the case where only magnetic charges are turned on.
The field strengths $F^a$ are $(d-2)$-forms.\footnote{
In particular, $*F^a \sim dt \wedge dr$.}
Again, we follow~\cite{hep-th/0507096} to obtain the key equations.
The ansatz for the metric, gauge field strengths, and scalars is
\bea
\mathrm{d}s^2 &=& -a(r)^2\, dt^2 + a(r)^{-2}\, dr^2 + b(r)^2\, \mathrm{d}\Omega_{d-2}^2 ~, \label{met_anz}\\
F^a &=& Q^a_m \sin^{d-3}\theta\, \sin^{d-4}\varphi\, \ldots\, d\theta \wedge d\varphi \wedge \ldots ~, \label{gft_anz} \\
\phi^i &=& \phi^i(r) \label{phi_anz} ~.
\eea
The equations of motion analogous to~\eref{eq:eom1}--\eref{eq:eom3} then tell us that
\bea
&& -2b^2V_g  
= (d-3)^2(-1+a^2b^{\prime\,2})+\tfrac{1}{2}b^2(a^2)'' + ab((3d-8)a'b'+(d-3)ab'')  ~, \label{eq:neweom1} \\
&& \frac{(d-2)b''}{2b} = -\sum_i \big(\phi^{i\, \prime}\big)^2 ~, \label{eq:neweom2} \\
&& \left(a^2b^{d-2}\phi^{i\,\prime}\right)' = \frac{(d-2)!\partial_i V_\mathrm{eff}(\phi)}{4b^{d-2}} +\tfrac{1}{2}b^{d-2}\p_iV_g~, \label{eq:neweom3}
\eea
where the effective potential is
\be
V_\mathrm{eff}(\phi) = f_{ab}(\phi) Q^a_m Q^b_m ~. \label{eq:newveff}
\ee
The Hamiltonian constraint becomes
\be
-(d-2)[(d-3)-ab'(2a'b+(d-3)ab')] = 2\sum_i \big(\phi^{i\, \prime}\big)^2 a^2b^2-\frac{(d-2)!V_\mathrm{eff}(\phi)}{b^{2(d-3)}} - 2 b^2 V_g(\phi) ~. \label{eq:newhc}
\ee

\subsection{Change of coordinates}
Using the variable transformation
\be
\partial_r = (d-3) b^{d-4}\partial_\rho ~, \label{eq:rho}
\ee
we can put the equations of motion into a form similar to those in the preceding section.
Note that the units of $\rho$ are $  [L]^{d-3}$.
In these new coordinates, in analogy to~\eref{eq:eom1}, the relation~\eref{eq:neweom1} becomes
\be
\p_\rho^2(a^2 b^{2(d-3)})=2 -\frac{4 b^2}{(d-3)^2} V_g ~. \label{eq:hd_1}
\ee
For the other equations of motion from~\cite{hep-th/0507096} using~\eref{eq:neweom2}, one finds that~\eref{eq:newhc} can be written as
\begin{equation}
-\tfrac{1}{2}\p_\rho \left( a^2 \p_\rho (b^{2(d-3)})\right)   = -1 + (d-4)!\frac{ V_{\mathrm{eff}}(\phi)}{b^{2(d-3)}} + \frac{2b^2}{(d-2)(d-3)}V_g(\phi)~,
\label{ham_hd}
\end{equation}
and the equation of motion for the scalars~\eref{eq:neweom3} becomes
\begin{equation}
\p_{\rho} \left(a^2b^{2(d-3)}\p_\rho\phi^i\right)=\frac{(d-2)!}{4(d-3)^2}
  \frac{\partial_i V_\mathrm{eff}(\phi)} {b^{2(d-3)}}+ \frac{1}{2(d-3)^2}b^2\p_iV_g~.
  \label{dilatoneqhd}
\end{equation}
Finally,~\eref{eq:neweom2} becomes
\begin{equation}
\sum_i (\p_\rho \phi^i)^2= -\frac{d-2}{2(d-3)}\frac{\p_\rho^2(b^{d-3})}{b^{d-3}}
\label{eq:bdashdash}
\end{equation}
It is straightforward to check that for $d=4$, we recover the equations from before.

If we have an extremal horizon  with $a=a'=0$, evaluating \eqref{eq:neweom3}  at the horizon yields
\begin{equation}
\label{gen_atr_d}
\frac{(d-2)!\partial_i V_\mathrm{eff}(\phi)}{2b_\text{ext}^{2(d-3)}}+b_\text{ext}^2\p_iV_g(\phi)=0 ~.
\end{equation}
As for the lower dimensional case,  we restrict our attention to solutions to \eqref{gen_atr_d} which satisfy \eqref{min}.
Then, evaluating the Hamiltonian constraint \eqref{eq:newhc} at the horizon yields
\begin{equation}
\label{gen_atr2_d}
 (d-4)! V_\mathrm{eff,min}  - \frac{2b_\text{ext}^{2(d-2)}}{(d-2)(d-3)}(-V_\text{$g$,min})-b_\text{ext}^{2(d-3)}=0 ~.
\end{equation}
Given $V_\mathrm{eff,min}>0$ and $V_\text{$g$,min}<0$, the polynomial in~\eref{gen_atr2_d} crosses the $b_\mathrm{ext}^2$ axis exactly once.
There is thus a unique positive real valued solution for $b_\text{ext}$ which is solely a function of $V_\mathrm{eff,min}$ and $V_\text{$g$,min}$.

When we do not have an extremal horizon, we may integrate~\eqref{ham_hd} and~\eqref{dilatoneqhd} between two horizons $\rho_\pm$ (which correspond to zeros of $a$) to give the averaged attractor equations
\begin{eqnarray}
\int^{\rho_+}_{\rho_-}d\rho\ \left(\frac{(d-2)!\partial_i V_\mathrm{eff}(\phi)}{2b^{2(d-3)}}+b^2\p_iV_g(\phi)\right) &=& 0 ~,\label{hot1}\\
\int^{\rho_+}_{\rho_-}d\rho\ \left(\frac{ (d-4)! V_{\mathrm{eff}}(\phi)}{b^{2(d-3)}} + \frac{2b^2}{(d-2)(d-3)}V_g(\phi)-1  \right)&=&0 ~.\label{hot2}
\end{eqnarray}
Note that at zero temperature the integrands of the $d$ dimensional attractor equations vanish on their own.

\subsubsection{$V_g = 0$ and a Reissner--Nordstr\"om-like solution}
When $V_g=0$, integrating~\eqref{eq:hd_1} yields
\be
a^2 b^{2(d-3)}=(\rho-\rho_+)(\rho-\rho_-) ~. \label{eq:ab_hd}
\ee
From~\eqref{eq:ab_hd} we see that, assuming that the horizon area is non-zero, $a^2$ has two possible roots corresponding to horizons at $\rho_+$ and $\rho_-$.

If we assume constant scalars, $\phi_0$ that satisfy  $\partial_i V_\mathrm{eff}(\phi)=0$, which is to say, we fix the scalar fields to their attractor values, we can construct Reissner--Nordstr\"om-like solutions. 
Substituting the ansatz $b_0=r$ in~\eqref{eq:ab_hd} and using~\eref{eq:rho}, we determine
\begin{equation}
a_0^2=\frac{(\rho-\rho_+)(\rho-\rho_-)}{\rho^2} = \frac{(r^{d-3}-r^{d-3}_+)(r^{d-3}-r^{d-3}_-)}{r^{2(d-3)}} ~,
\end{equation}
where $\rho_\pm$ and $r_\pm$ are integration constants corresponding to the positions of the horizons.
Consistent with the Hamiltonian constraint~\eref{ham_hd}, we find
\be
V_\mathrm{eff}(\phi_0) = \frac{\rho_+\rho_-}{(d-4)!} = \frac{\rho_\mathrm{ext}^2}{(d-4)!} ~.
\ee
This is in accord with the area law.
The relation $A_+ A_- = A_\mathrm{ext}^2$ follows immediately from $\rho_+ \rho_- = \rho_\mathrm{ext}^2$, which we have derived above.

The near horizon geometry of the extremal Reissner--Nordstr\"om solution is AdS$_2\times S^{d-2}$ and is a solution of supergravity in its own right.
In the non-extremal case, a Harrison transformation~\cite{cl1,cl2,cg} enables us to translate the flat space Reissner--Nordstr\"om solution to a black hole with AdS$_2\times S^{d-2}$ asymptopia.
Using the graviphoton, we can then uplift AdS$_2$ to obtain a BTZ black hole in AdS$_3$ with the same non-extremality parameter and identical thermodynamic properties of temperature and entropy.
The Cardy formula in the dual CFT$_2$ enables us to calculate the entropy~\cite{gjn2}.
Since the generator of time translations is the $L_0 + \overline{L}_0$ combination of Virasoro generators, we can change the mass while leaving the orthogonal combination $L_0 - \overline{L}_0$, which is the generator of spatial translations corresponding to momentum along a circle, unchanged.
This explains why the $J = L_0-\overline{L}_0 \propto r_+ r_-$ eigenvalue is independent of the non-extremality parameter.
This provides a CFT interpretation of the area law~\cite{gjn1}.

\subsubsection{$V_g\neq 0$, constant scalars}\label{subsec:vgnonzero}
For $V_g\neq 0$ but with constant scalars, \eqref{eq:bdashdash} together with \eqref{eq:rho} implies that
\begin{equation}
\label{eq:bsol}
b^{d-3}=\rho = r^{d-3} ~.
\end{equation}
Now, as for the lower dimensional case, given that the right hand side of \eqref{dilatoneqhd} must be constant and the two terms on the left hand side have different $r$ dependence we again find \eqref{min} holds.
Then \eqref{eq:ab_hd} becomes 
\be
\p_\rho^2(a^2 \rho^{2})=2 -\frac{4}{(d-3)^2} \rho^{\frac{2}{d-3}} V_{g,\text{min}} ~. \label{eq:hd_2}
\ee
The solution written in terms of $r$ is
\begin{equation}
\label{eq:ab_sol}
a^2b^{2(d-3)}=\frac{2 (-V_{g,\text{min}})}{(d-1)(d-2)}r^{2(d-2)}+r^{2(d-3)}-c_1r^{d-3}+c_2~,
\end{equation}
where once again, negativity of $g_{tt}$ for large $r$ requires $V_{g,\text{min}}<0$.
From the large $r$ limit of \eqref{eq:ab_sol}, we see \cite{Chamblin:1999hg} that
\begin{eqnarray}
\label{c1M}
c_1 &=& \frac{16\pi G_d M}{(d-2)\omega_{d-2}}~, \\
V_{g,\text{min}}&=&\Lambda_\text{eff}~, \label{eq:lameff}
\end{eqnarray}
where $\omega_{d-2}$ is the volume of a unit $(d-2)$ sphere, $M$ is the ADM mass of the solution, and $\Lambda_\text{eff}$ is the effective cosmological constant.
Using the Hamiltonian constraint \eqref{eq:newhc}, one finds that
\begin{equation}
\label{c2_gen}
c_2 = (d-4)!\, V_\text{eff,min}~.
\end{equation}
With $V_{g,\text{min}}$ negative and $M$ positive, when $d>3$, we find that \eqref{eq:ab_sol} has at most two real strictly positive roots.\footnote{
Calculating the Ricci scalar given \eqref{met_anz}, one finds $R\sim\frac{1}{b^2}$. So for the constant scalar case, where $b=r$, there is a curvature singularity at $r=0$.  This means that we are only interested in positive roots of \eqref{eq:ab_sol}.}
\begin{figure}[htb]
	\centering
	\includegraphics[width=0.5\linewidth]{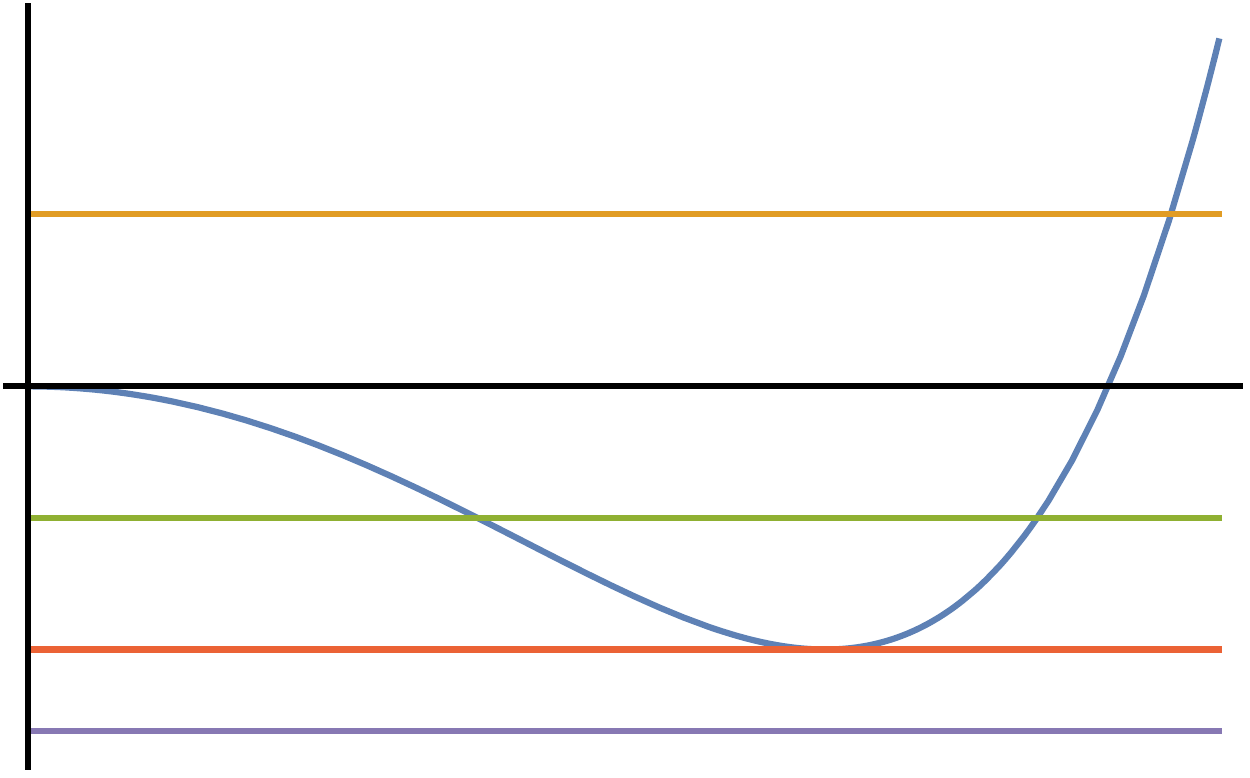}
	\caption{Rough plot of \eqref{eq:ab_sol}. One sees that, as we vary $c_2$, we can have $0$, $1$ or $2$ strictly positive roots.}		
	\label{fig:ab}
\end{figure} 
To see this notice that, when $d>3$, for $r$ sufficiently small, the third term of \eqref{eq:ab_sol} dominates and, since the term has a negative coefficient, the polynomial is decreasing near $r=0$. As $r$ increases, the first two terms will dominate so that the polynomial will eventually curve upwards and keep increasing. We conclude that, depending on the value of $c_2$, which can shift the curve up or down, \eqref{eq:ab_sol} can have at most two real strictly positive roots\footnote{
We see from Figure~\ref{fig:ab} that when $c_2$ is negative, there is a single positive root. This would correspond to the gauge fields having imaginary charges and negative energy density which we can discount as unphysical.   } --- see Figure~\ref{fig:ab}.

\subsection{Perturbation about the attractor point}\label{subsec:pert}
We perturb about the zero temperature Reissner--Nordstr\"om attractor solution (with $\rho_\pm=\rho_\mathrm{ext}$):
\begin{equation}
\phi^i = \phi^i_0 + \epsilon \phi^i_1 ~.
\end{equation}
The first order correction will satisfy
\begin{equation}
\p_{\rho} \left(a_0^2b_0^{2(d-3)}\p_\rho(\phi^i_1)\right)= m_i^2 \frac{\phi^i_1} {b_0^{2(d-3)}} ~, \qquad
m_i^2 = \frac{(d-2)!}{4(d-3)^2} \partial_i^2 V_\mathrm{eff}(\phi_0) ~,
  \label{dilatoneqhd1}
\end{equation}
where $m_i^2$ the coefficient of the first order expansion of the right hand side of \eqref{dilatoneqhd}.
This has the solutions
\begin{equation}
\phi^i_1=c^i_1 \left(1-\frac{\rho_\mathrm{ext}}{\rho}\right)^{\gamma_i} ~,
\end{equation}
with 
\begin{equation}
\gamma_i=\frac{1}{2}\left(\pm\sqrt{1+\tfrac{4m_i^2}{\rho_\mathrm{ext}^2}}-1\right)~.
\label{eq:gammai}
\end{equation}
In order to have a solution that does not blow up on the horizon, we take the positive root.

A change of coordinates
\be
z = \frac{\rho_+(\rho-\rho_-)+\rho_-(\rho-\rho_+)}{\rho(\rho_+-\rho_-)}
\ee
renders~\eref{dilatoneqhd1}
\be
\partial_z(\rho_\mathrm{ext}^2(z^2-1)\partial_z\phi^i_1) = m_i^2\phi^i_1 ~. \qquad
\ee
This is the Klein--Gordon equation for static scalar fluctuations with mass $m_i$ in AdS$_2$ with metric\footnote{
The quantities $\rho_\mathrm{ext}$ and $m_i$ are rescaled by powers of the AdS length $\ell$ in order to fix the appropriate units.}
\be
\mathrm{d}s^2 = -\rho_\mathrm{ext}^2(z^2-1) dt^2 + \frac{\ell^2\, dz^2}{\rho_\mathrm{ext}^2(z^2-1)} ~.
\ee
We therefore expect an underlying $SL(2,\mathbb{R})$ symmetry.
This is again suggestive of an AdS/CFT description of the black hole's entropy.

As known from~\cite{hep-th/0507096}, the first order correction to the scalar sources second order corrections to the metric.
We therefore write
\begin{eqnarray}
b^{d-3}&=&b_0^{d-3}+\epsilon^2 b_2 ~, \\
a^2 &=& a_0^2 +\epsilon^2 a_2 ~.
\end{eqnarray}
Expanding~\eqref{eq:ab_hd} to second order, we obtain
\begin{equation}
2a_0^2b_0^{d-3}b_2+b_0^{2(d-3)}a_2=0 ~,
\end{equation}
which implies
\begin{equation}
a_2 = -2\left(1-\frac{\rho_\mathrm{ext}}{\rho}\right)^2\frac{b_2}{\rho} ~.
\end{equation}
Using~\eqref{eq:bdashdash}, we compute
\begin{equation}
b_2 = \sum_i d^i_2 \rho \left(1-\frac{\rho_\mathrm{ext}}{\rho}\right)^{2\gamma_i} ~, 
\end{equation}
where 
\begin{equation}
d^i_2=-\frac{(c^i_1)^2(d-3)\gamma_i}{(d-2)(2\gamma_i-1)} ~.
\end{equation}

\subsection{Exact solution for a single scalar}\label{subsec:exact}
Again following~\cite{hep-th/0506177}, we aim to construct an exact solution for the scalars in arbitrary dimension.
For simplicity, we take a single scalar that couples to a pair of gauge fields with $V_g=0$ and consider a magnetically charged solution.
In this case, the effective potential has the form
\be
\label{veff}
V_\mathrm{eff}(\phi) = \sum_{a=1}^2 e^{\alpha_a\phi} (Q_m^a)^2 ~.
\ee
As shown in Appendix~\ref{ap:exact}, for the special case, 
\begin{equation}
\label{alpha_conf}
\alpha_1=-\alpha_2=\alpha = \sqrt\frac{8(d-3)}{d-2} ~,
\end{equation}
we obtain the solution
\begin{eqnarray}
e^{\alpha\phi} &=& e^{\alpha\phi_\infty} 
\frac{\left(\rho+\tfrac{1}{2}\alpha\Sigma \right)}{\left( \rho-\tfrac{1}{2}\alpha\Sigma \right)} ~, \label{eq:simpler2}\\
a^2 &=& \frac{(\rho-\rho_+)(\rho-\rho_-)}{b^{2(d-3)}} ~,\\
b^{2(d-3)}&=&\rho^2-\tfrac{1}{4}\alpha^2\Sigma^2 ~, \label{b_simpler}
\end{eqnarray}
where $\Sigma$ is the so called scalar charge defined from the expansion
\begin{equation}
\label{eq:sigma_def}
\phi=\phi_\infty + \frac{\Sigma}{r^{d-3}}+\ldots = \phi_\infty + \frac{\Sigma}{\rho}+\ldots ~.
\end{equation}
In addition, we can read off mass of the solution from the asymptotic form of $a^2$, obtaining
\begin{equation}
\label{Mrho}
M=\frac{\rho_++\rho_-}{2} ~.
\end{equation}
It is also worth noting that the scalar charge is not an independent parameter:
\begin{equation}
\label{sig}
\Sigma = \frac{ (d-4)!((\bar{Q}_m^2)^2-(\bar{Q}_m^1)^2)}{\alpha M} ~,
\end{equation}
where
\begin{equation}
\label{Qdef}
\bar Q_m^a = e^{\frac{1}{2}\alpha_i \phi_\infty}Q_m^a~.
\end{equation}

In order to determine the relationship between $\rho$ and the radial coordinate, $r$, it is convenient to define $\scripty{r} =(\rho-\tfrac{1}{2}\alpha |\Sigma|)/\alpha|\Sigma| $ so that
\begin{equation}
\label{b_scaled}
b=|\alpha \Sigma|^{\frac{1}{d-3}}(\scripty{r}(\scripty{r}+1))^{\frac{1}{2(d-3)}}
\end{equation}
and
\begin{equation}
\label{phi_nic}
e^{\alpha\phi}=e^{\alpha\phi_\infty}\left(1+\frac{1}{\scripty{r}}\right)^{\text{sign}(\Sigma)}
\end{equation}
Integrating~\eqref{eq:rho}, we finally obtain
\begin{equation}
\label{label}
r = \frac{|\alpha\Sigma|}{d-3}\int_0^\scripty{r} d\bar{\scripty{r}}\ b^{4-d}(\bar{\scripty{r}}) 	
=\frac{|\alpha \Sigma|^{\frac{1}{d-3}} }{\tfrac{1}{2}(d-2)}
{ \scripty{r}^{\left(\frac{1}{2}+\frac{1}{2 (d-3)}\right)} \, _2F_1\left(\tfrac{1}{2}-\tfrac{1}{2 (d-3)},\tfrac{1}{2}+\tfrac{1}{2 (d-3)}; \tfrac{3}{2}+\tfrac{1}{2(d-3)};-\scripty{r}\right)}
~,
\end{equation}
where the lower limit of integration is chosen so that $b(r=0)=0$.

In the extremal limit, we find
\begin{eqnarray}
\label{sigma_ex}
\tfrac{1}{2}\alpha\Sigma &=& \sqrt{\frac{(d-4)!}{2}}\left(\bar Q_m^2-\bar Q_m^1\right) ~, \\
\label{m_x}
M &=&  \sqrt{\frac{(d-4)!}{2}}\left(\bar Q_m^2+\bar Q_m^1\right) ~.
\end{eqnarray}

\section{Attractors in infinite dimensions}\label{sec:infd}

\begin{figure}[htb]
	\centering
	\includegraphics[width=\linewidth]{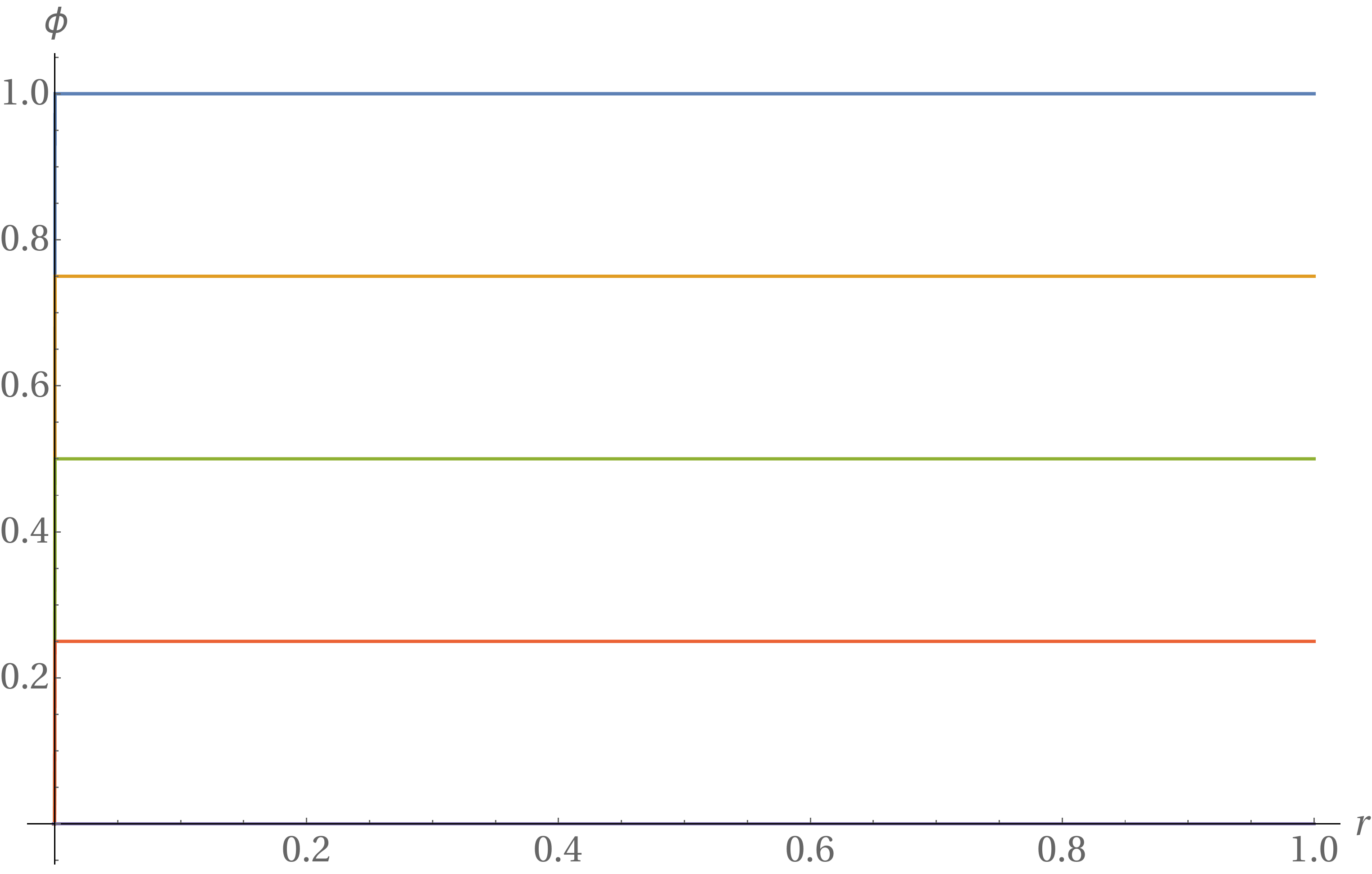}
	\caption{Cartoon of the qualitative attractor behavior we expect in infinite dimensions. The fields remain constant until one reaches the horizon where they suddenly jump to their attractor value.}
	\label{fig:cartoon}
\end{figure}

\begin{figure}[htb]
	\centering
	\includegraphics[width=\linewidth]{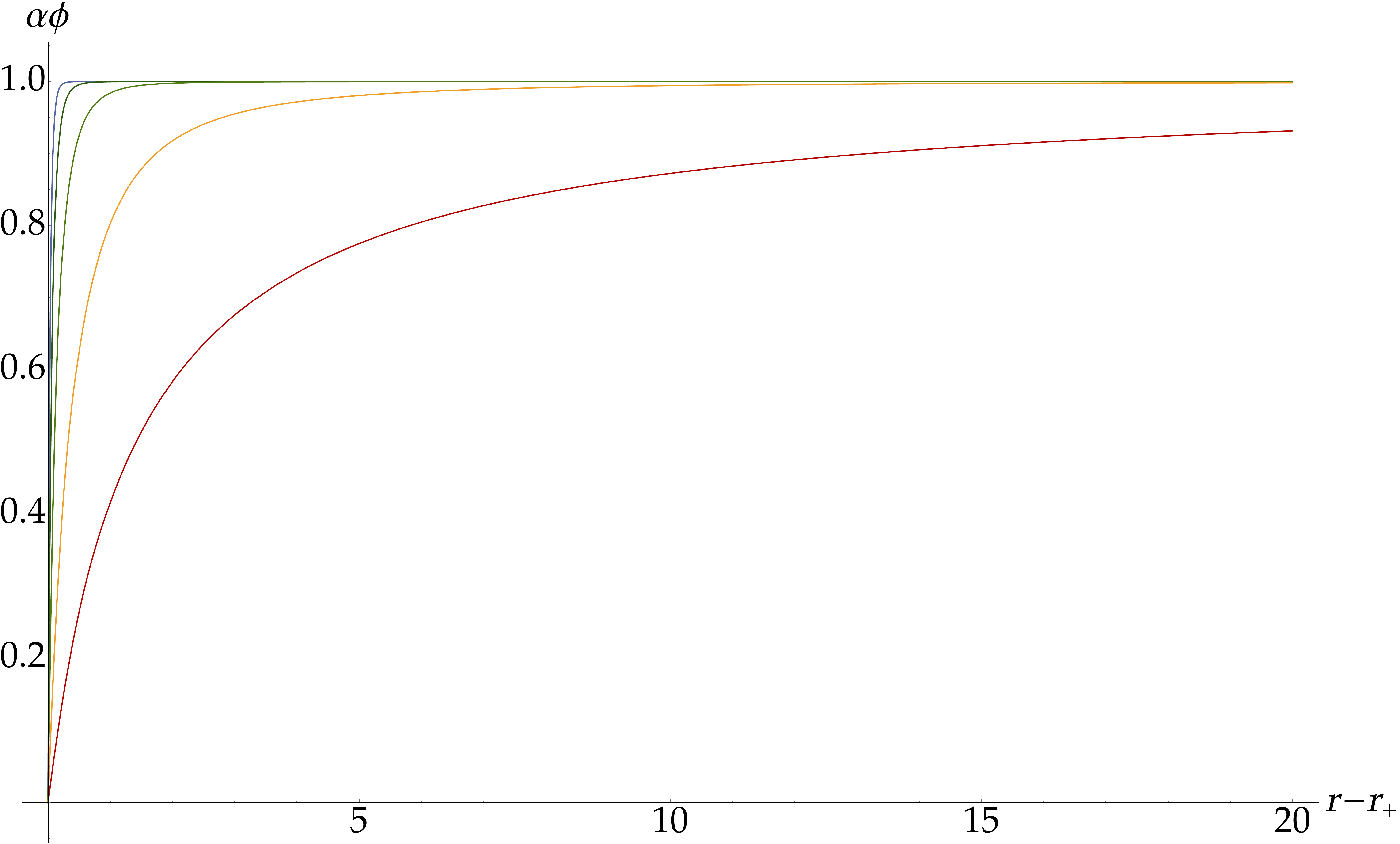}
	\caption{Plot of the solution, \eqref{eq:simpler2}, in the extremal limit, for $d=4,5,7,10,14$ (with $Q_1=Q_2=1$ and $\alpha\phi_\infty=1$).
		For these parameters, $\phi_{0}=0$. We have shifted the radial coordinate so that the horizon is at $r=0$.}
	\label{fig:phiD}
\end{figure}

In a $d$ dimensional spacetime,  we expect the gravitational Newtonian potential of a point source to fall off like $r^{3-d}$.  
In the $d\rightarrow\infty$ limit, the effective range of the gravitational force approaches $0$.  
As discussed in \cite{1302.6382}, this feature extends to general relativity. It follows that if the attractor mechanism still holds as   
$d\rightarrow\infty$, one suspects that scalars only start flowing to their attractor value very near the horizon. This suggests that  in infinite dimensions, as illustrated in Figure~\ref{fig:cartoon}, away from the horizon, scalars would not flow, maintaining their asymptotic value and suddenly jumping to the attractor value at the horizon.\footnote{
More precisely, following \cite{1302.6382}, we expect the ``jump'' to occur over a length scale $1/d$ times smaller than the horizon scale.}
This behavior is confirmed by studying the exact extremal solutions found in the previous section, where as illustrated in Figure~\ref{fig:phiD}, we see that for a given asymptotic value the flow approaches a step function as we increase the number of spatial dimensions.\footnote{
We have slightly cheated in plotting Figure~\ref{fig:phiD} in that the value of $\alpha$ used to make the plot itself depends on $d$. We expect the qualitative features of the graph will remain the same if we keep $\alpha$ constant while increasing $d$.}

\section{Horizon invariants}\label{sec:cplx}

In this section we discuss invariants that can be formed from the products of horizon areas.

\subsection{Higher dimensional area law}\label{sec:hdal}

When $V_g=0$, using the results of Section~\ref{sec:gd}, it is straightforward to generalize the proof of the area law in \cite{gjn1}.
We calculate the temperatures from computing the periodicity of the Euclidean time direction at two real horizons.
This gives
\begin{eqnarray}
T_{\pm} = \frac{(a^2)'_\pm}{4\pi} ~,\label{6}
\end{eqnarray}
where the $+$ and $-$ subscripts denote the outer and inner horizons, respectively.
We then employ~\eref{eq:rho} and~\eref{eq:ab_hd} to obtain the following expressions  for the temperatures:
\begin{equation}
4\pi T_\pm=(a^2)'_\pm = \pm \frac{2(d-3)\Delta}{b^{d-2}_\pm} ~, \label{7}
\end{equation}
where $\Delta = \tfrac{1}{2}({\rho_+  -  \rho_-})$ measures the non-extremality, which, as can be seen from (\ref{7}), goes to zero in the extremal case and is directly proportional to the temperatures, $T_\pm$.
Notice that the temperature of the inner horizon is negative.
This is simply the statement that the area of the inner horizon becomes smaller as the non-extremality parameter $\Delta$ increases.\footnote{
	In the BTZ black hole, $T_\pm^{-1} = \frac12 (T_R^{-1} \pm T_L^{-1})$.
	The temperatures of the left and right moving sectors of the dual CFT$_2$ are always positive.
	There is no pathology inherent to associating a negative temperature to the inner horizon in the bulk.
	Inner horizons do suffer classical instabilities~\cite{penrose68,poisson90}.
	In this work, we do not consider dynamical properties of the inner horizon.}
When we interpret area as entropy in accordance with the first law of black hole mechanics, this means the inner horizon encodes fewer degrees of freedom the more non-extremal the solution gets.
As the area of the $S^{d-2}$ horizon scales like $b^{d-2}$, from~\eref{7} we have
\begin{equation}
A_+ T_+ + A_- T_- = 0 ~. \label{AT_rel}
\end{equation}
If the entropy scales with the area, we may use~\eref{AT_rel} to readily deduce that
\be
\left. \frac{\partial}{\partial M} S_+ S_- \right|_{Q^a_m} = \frac{S_+ T_+ + S_- T_-}{T_+ T_-} = 0 ~.
\ee
From the extremal solution, the product of the entropies and therefore the areas is mass independent:\footnote{
	We have taken a shortcut by assuming that $S_\pm \propto A_\pm$.
	This simplifies the derivation, but the assumption is not necessary.
	Using the arguments from Appendix~C of our prior work~\cite{gjn1}, we can obtain the area relation $A_+ A_- = A_\mathrm{ext}^2$ directly without invoking the proportionality of area and entropy.
	The area law is as well more general than the equivalent statement about entropy.
	Consistent with diffeomorphism invariance of the gravitational action, entropy may be computed as a Noether charge using the Wald formula.
	When $A_+A_- = A_\mathrm{ext}^2$,
	the inequality $S_+S_-\ne S_\mathrm{ext}^2$ signals a violation of the Smarr relation~\cite{cretc5}.}
\begin{equation}
\label{eq:mass_indep}
S_+S_-=S_{\mathrm{ext}}^2 \qquad \Longrightarrow \qquad A_+ A_- = A_\mathrm{ext}^2 ~.
\end{equation}

On the other hand, when $V_g\neq0$ we note that, for the constant scalar case in four dimensions, from  \eqref{a2b2:quartic} and \eqref{const}, 
the product of the positions of the horizons is 
\begin{equation}
\label{prod}
r_+r_- |\lambda|^2 = \frac{3 V_\mathrm{eff,min}  }{(-V_\text{$g$,min})  }~.
\end{equation}
Using the fact that for constant scalars $b(r)=r$, from \eqref{bext_gen},
\begin{equation}
\label{prod2}
r_\text{ext}^4= \left(\frac{\sqrt{1+4 V_\mathrm{eff,min}(-V_\text{$g$,min})}-1}{2(-V_\text{$g$,min})}\right)^2~.
\end{equation}
Comparing \eqref{prod} to \eqref{prod2} we see that, unlike the case when $V_g=0$, the geometric mean of the horizons areas is not, in general, equal to the extremal area.

\subsection{Only two real zeros}

Before we consider the general case, to illustrate our method, we initially set $V_g = 0$ and work in four dimensions and recover known results.
In the simple case, taking $r_1=r_-$ and $r_2=r_+$ to be the radial coordinates of the two horizons of a non-extremal spherically symmetric black hole and integrating \eqref{eq:eom1} 
between the horizons, we obtain the constraint, 
\begin{equation}
\sum^2_{i=1} (-)^i\left.(a^2)' b^2 \right|_{r=r_i} = 4 \Delta = 2(r_2-r_1)=\alpha(r_1, r_2) ~.
\end{equation}
In writing this expression, we have used the fact that $a^2(r_i) = 0$ and employed~\eref{7}.
Now, there exists $l^-$ and $l^+$, with $l^+ - l^- =1$, such that
\begin{eqnarray}
\label{eq:lpm}
\left.(a^2)' b^2 \right|_{r=r_1}  &=& \alpha\, l^-~, \cr
\left.(a^2)' b^2 \right|_{r=r_2}  &=& \alpha\, l^+~. 
\end{eqnarray}
On-shell static spherically symmetric configurations extremize the following effective one dimensional action:
\begin{equation}
S_{1d} = \int dr \left((a^2 b)'b'- a^2 b^2 (\phi')^2- \frac{V_\mathrm{eff}(\phi)}{b^2} + \text{total derivative}\right)~.
\end{equation}
That is to say, the equations of motion~\eref{eq:neweom1}--\eref{eq:neweom3} are obtained from variation of this effective action.
The Palatini variation of the action is performed by imposing constraints that preserve the asymptotic boundary conditions:
\begin{equation}
\label{eq:a_asymp}
a^2 = 1- \frac{2 M}{r}+\ldots,
\end{equation} 
\begin{equation} 
\label{eq:b_asymp}
b^2 = r^2 + \text{constant}+\ldots,
\end{equation} and  
\begin{equation} 
\label{eq:f_asymp}
\phi(r) = \phi_{\infty} + \frac{\Sigma}{r}+\ldots\;.
\end{equation} 
We must also impose the Hamiltonian constraint~\eref{eq:newhc}.
The functions $a(r)$, $b(r)$ and $\phi(r)$ that satisfy these equations define our solution space.
In particular, we can parametrize the solution space by coordinates $(a^2)'_\pm$ and the asymptotic values of the scalar fields.

We perform the variation only on the on-shell field configurations.
The volume term arising from a variation about the extremum  vanishes, and we are left with total derivative or surface terms. Schematically, for some degree of freedom $\Phi$, after varying the action and performing integration by parts on the term $\frac{\delta L}{\delta (\partial \Phi)} \delta (\partial \Phi$) we obtain a surface term:
\begin{equation}
\label{eq:var}
\int_{r_1}^{r_2}dr\left(\frac{\delta L}{\delta (\partial \Phi)} \delta \Phi\right)'=\left.\frac{\delta L}{\delta (\partial \Phi)} \delta \Phi\right|_{r_1}^{r_2}=0~.
\end{equation}
 Note that this surface term receives no contribution from the potential terms, $ \frac{\partial L}{\partial  \Phi}$, but only from kinetic terms containing $\partial \Phi$.\footnote{
 Note that these surface terms arise from an Euler--Lagrange variation of the bulk action and are independent of arbitrary surface terms that can be added to the action and which do not feed into the Euler--Lagrange variation.} In our case, we have metric degrees of freedom $a^2$ and $b^2$ and the scalar fields coupled to gravity. 

 We now consider variations of the effective action evaluated with boundaries at the event and Cauchy horizons. On any given horizon, $a^2=0$, so $\delta a^2 =0$.
 This tells us that
 \begin{equation}
 \label{eq:var2}
 \left.\frac{\delta L}{\delta (\partial \Phi)} \delta \Phi\right|_{r_1}^{r_2} = 
 \left[\frac{\delta L}{\delta ((b^2)')} \delta b^2+\frac{\delta L}{\delta ((\phi^i)')} \delta \phi^i \right]_{r_1}^{r_2} = (a^2)' \delta b^2 |^{r_2}_{r_1}=0~,
 \end{equation}
where we have evaluated ${\delta L}/{\delta ((b^2)')}$ and used the fact that  
 the remaining surface terms proportional to ${\delta L}/{\delta ((\phi^i)')}$ contain factors of $a^2 b^2$, which vanish at the boundaries. (See Appendix C of~\cite{gjn1} for details.)
 Using~\eqref{eq:lpm}, we can recast~\eqref{eq:var2} as 
\begin{equation}\label{c0}
\alpha\, (l^+ \delta  ( \log\, b_2^2) -l^- \delta (\log \, b_1^2)) = 0~,
\end{equation}
where $b^2_i = b^2(r_i)$, which in turn implies 
\begin{equation}\label{c}
l^+ \delta  ( \log\, b_2^2) =l^- \delta (\log \, b_1^2)~.
\end{equation}
The above variation is performed in the space of on-shell solutions,  at fixed values of charges, collectively denoted by $Q$, and as such, every point in this space is characterized by the value of the coupling constants at asymptotic radial infinity. 
We are keeping the gauge charges constant because this is what makes sense in the context of the attractor mechanism.
For extremal solutions, the scalars flow to the minimum of the effective potential and assume attractor values that are functions of the charges.
We wish to preserve a smooth zero temperature limit in which we recover this behavior.
The hot attractor mechanism therefore focuses on solutions with fixed charges.\footnote{
As a parallel investigation, we could keep the mass fixed and allow the gauge charges to vary.
While this may be an interesting exercise in its own right, it is not connected to the generalization of the attractor mechanism that we consider in this paper.}

Now consider an allowed variation from a fiducial point in solution space labeled $*$ to another point in solution space labeled $**$, with $\log b^2_{**}=\log b^2_*+\delta \left(\log b^2\right)$. The aforementioned constraint, \eqref{c}, will take the form, 
\begin{equation}\label{c1}
l^+_* \delta  ( \log\, b_2^2) =l^-_* \delta (\log \, b_1^2)~.
\end{equation}
The reverse variation yields
\begin{equation}\label{c2}
l^+_{**} [-\delta  ( \log\, b_2^2)] =l^-_{**}[- \delta (\log \, b_1^2)]~.
\end{equation}
Taking the ratio  of \eqref{c1} and \eqref{c2}, we have
\begin{equation}\label{c3}
 \frac{l^+_{**}}{l^+_*}  -\frac{l^-_{**}}{l^-_*} = 0~.
\end{equation}
Since this is true for any pair of points in solution space that can be connected by an infinitesimal transformation,  assuming that the solution space is connected, we see that at any point, the ratio $\frac{l^-}{l^+}$ is a constant. 
Therefore the generic boundary variation constraint \eqref{c} can be written 
\begin{equation}
\delta \left( \log (b^{2 }(r_1) b^{-2 (l^-/l^+)}(r_2)) \right) = 0
\end{equation}
implying that 
\begin{equation}\label{d}
b(r_1) b^{-(l^-/l^+)}(r_2)=F(Q) ~.
\end{equation}
Notice again that variations over the space of solutions have kept the charges fixed.
The right hand side therefore is a constant.

To determine what this constant is,
we further assume that there exists a point in solution space characterized by purely constant scalars.
As we have seen previously, at this point $b(r) = r$.
The metric in this case is always Reissner--Nordstr\"om with the property that
\begin{equation}
r_+ r_- = F(Q) = r_\mathrm{ext}^2 ~.
\end{equation}
Comparing this with \eqref{d}, we see that $l^+ = -l^-$, and hence the invariant constraint on the space of solutions is
\begin{equation}
\delta \left( b^{2}(r_1) b^{2}(r_2) \right) = 0 ~.
\end{equation}
This derives the area law:
\be
(4\pi b^2(r_1)) (4\pi b^2(r_2)) = A_+ A_- = A_\mathrm{ext}^2 ~.
\ee

\subsection{$N$ complex zeros}
We extend this technique to the case where $a(r)$ has $N$ zeros. Let the position of the zeros of $a$ be $r_i$, where $1\leqslant i \leqslant N$.
 There is now a boundary for the variational problem at each $r_i$. Formally we take $r$ to be a complex variable.
 Let us consider the effective lower dimensional action, 
 \begin{equation}
 \label{aohdv}
 \begin{array}{cccl}
 S_{1d}&=&\int dr &
 \Big((d-3)(d-2)b^{d-4}(1+a^{2}b^{'2})+(d-2)b^{d-3}(a^{2})^{'}b^{'} \\
 &&&   -2a^{2}b^{d-2}(\partial_{r
 }\phi)^{2}-\frac{(d-2)!}{b^{d-2}}V_\text{eff}- 2 b^{d-2}V_{g}\Big)\\
 &=& \int d\rho &\Big( (d-2)\left(1 + a^2(\p_{\rho} b^{d-3})^2+\tfrac{1}{2}\p_\rho(a^2)\p_\rho[(b^{d-3})^2]\right)\\
 &&&   -2(d-3)a^2(b^{d-3})^2(\partial_{\rho}\phi)^{2}-\frac{(d-2)!}{(d-3)(b^{d-3})^2}V_\text{eff}- \frac{2}{(d-3)} b^{2}V_{g}\Big)
 \end{array}
 \end{equation}
  as a line integral along a contour ${\cal C}$ in the complex $\rho$-plane.
  This one dimensional effective action is obtained from integrating out the gauge charges from the equations of motion.
Unless all of the roots lie along a line, we can always form a simple polygon on the complex plane with the roots at the vertices.
We define ${\cal C}$ as this $N$-gon and order the vertices $\rho_1, \ldots, \rho_N$. 

Suppose we consider the variation of an on-shell configuration.
The variation remains on-shell along any real line.
(This line may connect, for example, the horizon to the asymptotic boundary.)
Along the real line the Euler--Lagrange equation for a degree of freedom $\Phi$ is
\begin{equation}
\label{euler}
\frac{\p L}{\p \Phi} - \frac{d}{d\rho}\left(\frac{\p L}{\p (\p_{\rho}\Phi)}\right)=0 ~.
\end{equation}
Here, we wish to vary the configuration evaluated on the contour ${\cal C}$ in the complex plane.
On the $i$-th edge of the $N$-gon, the Euler--Lagrange equation becomes
\begin{equation}
\label{euler2}
\frac{\p L}{\p \Phi} - \frac{ds_i}{d\rho}\frac{d}{ds_i}\left(\frac{\p L}{\p \left(\frac{ds_i}{d\rho}\frac{\p\Phi}{\p s_i }\right)}\right)=0 ~,
\end{equation}
where we have written $\rho=\rho(s_i)$.
So long as $s_i$ is an affine parameter, \textit{i.e.}, when $\frac{d\rho}{ds_i}$ is constant, the configuration remains on-shell.
To ensure that this happens, we put
\be
\rho(s_i)=\rho_i +(\rho_{i+1}-\rho_i)s_i ~, \qquad 0\le s_i\le 1 ~,
\ee
to linearly interpolate between adjacent roots.

For on-shell configurations along ${\cal C}$, following the arguments of the previous subsection, integrating along the $i$-th edge gives
\begin{equation}
\label{3}
 \left.(a^2)' \delta b^{2(d-3)} \right|^{\rho_{i+1}}_{\rho_i}=\left.\frac{1}{D_i}\frac{d}{ds_i}(a^2) \delta b^{2(d-3)} \right|^{\rho_{i+1}}_{\rho_i}=0~,
\end{equation}
where $D_i=\frac{d\rho}{ds_i}=(\rho_{i+1}-\rho_i)$. 
Furthermore, upon integration along the 
 $i$-th edge, the equation of motion 
\begin{equation}
\p_{\rho}^2(a^2 b^{2(d-3)})= 2 - \frac{4}{(d-3)^2} V_g b^2
\end{equation}
yields
\begin{equation}\label{1}
 \left.\frac{1}{D_i}\left(\frac{d(a^2)}{ds_i}\right)  b^{2(d-3)} \right|^{\rho_{i+1}}_{\rho_i}\,= \alpha_i(\rho_i,\rho_{i+1})~.
\end{equation}
This implies that there exists $2N$ real numbers, $l^\pm_i$, with
\begin{equation}\label{2}
\alpha_i l^+_{i}= \left.\frac{1}{D_i}\left(\frac{d(a^2)}{ds_i}\right)  b^{2(d-3)} \right|_{\rho_i} ~, \qquad \alpha_i l^-_{i}=\left.\frac{1}{D_i}\left(\frac{d(a^2)}{ds_i}\right)  b^{2(d-3)} \right|_{\rho_{i+1}} ~, 
\end{equation} 
such that, $l^+_{i}-l_{i}^- =1$. 
Substituting \eqref{2} into \eqref{3}, we get
\begin{equation}
\label{2b}
\alpha_i\left( l^+_i \delta \log b_{i+1}^{2(d-3)} -l^-_i \delta \log b_{i}^{2(d-3)}\right)=0~.
\end{equation}
Following the argument in the two zero case, we then conclude that $l^-_i/l^+_i$ is a constant, so that \eqref{2b} can be written as
\begin{equation}
\label{2c}
\delta \left( \log ( b_{i+1}^{2(d-3)} b_{i}^{2(d-3)(-l^-_i/l^+_i)}) \right)=0~.
\end{equation}
Adding together the contributions \eqref{2c} for all the edges as we traverse the contour, we obtain
\begin{equation}
\label{2d}
\delta \left(\log \prod_{i=1}^{N} b_i^{2(d-3)\left(1-(l^-_i/l^+_i)\right)}\right)=0 ~.
\end{equation}
This means that on the space of solutions,
\be
\prod_{i=1}^{N} b_i^{1-(l^-_i/l^+_i)} =\mathrm{constant} ~.
\label{eq:constant}
\ee
Again, we emphasize that in obtaining this equation, we have considered variations in the space of solutions for which the gauge charges are held fixed.
Going into this are the assumptions that the solution space is continuous and is characterized by the temperatures at each of the horizons and the moduli at infinity.
A special point in the space of solutions is the extremal black hole spacetime.
Here, the hot attractor mechanism we are studying reduces to the usual attractor mechanism.
The temperature vanishes at the extremal horizon and the scalars flow from infinity to attractor values.
From our knowledge of what happens at the extremal point in solution space, we may deduce the constant on the right hand side.

To derive an expression for the right hand side of~\eref{eq:constant}, let us consider the point in solution space where we set the moduli at infinity to their attractor values.
Here, as we have seen in Section~\ref{subsec:vgnonzero}, $a^2 b^{2(d-3)}$ is a degree $2(d-2)$ polynomial in $r$.
From (\ref{eq:ab_sol}) and (\ref{c2_gen}), Vieta's formula tells us that the product of the roots is
\begin{equation}
\prod^{2(d-2)}_{i=1} r_i = \frac{(d-1)!}{2(d-3)}\frac{  V_\mathrm{eff,min}}{(- V_{g,\mathrm{min}})}~.
\label{eq:blahblah1}
\end{equation}
The roots specify the locations of the horizons on the complex $r$-plane.
From~\eref{eq:blahblah1}, the product of the positions of the horizons depends only on $V_{g,\mathrm{min}}$ and $V_\mathrm{eff,min}$.
The potential $V_g$ for the scalars appears in the action~\eref{eq:ACTION} as a cosmological term.
We recall from~\eref{eq:newveff} that $V_\mathrm{eff}$ is a function of the scalars and the charges.
The right hand side of~\eref{eq:blahblah1} is therefore determined completely by the charges and the attractor values of the moduli.

Let us hold the moduli fixed at the attractor values and consider variations of the solution that change, for instance, the mass or the temperature of the black hole.
Under such variations, we have shown that
\be
\delta (\prod^{2(d-2)}_{i=1} r_i) = \delta (\prod^{2(d-2)}_{i=1} b_i) =0 ~,
\ee
where the first equality uses the identification of coordinates in~\eref{eq:bsol}.
Comparing to~\eref{eq:constant}, we conclude from this that $(l^-_i/l^+_i) = (l^-_j/l^+_j)$, for all $i$ and $j$.
Hence, analysis of the solution at the special point in moduli space where the scalars assume the attractor values tells us that the general on-shell $2(d-2)$-horizon solution space has an invariant defined by 
$\delta ( \prod^{2(d-2)}_{i=1} b_i ) = 0$
and this invariant is
\begin{equation}
\prod^{2(d-2)}_{i=1} 	b_i = \frac{(d-1)!}{2(d-3)}\frac{  V_\mathrm{eff,min}}{(- V_\text{$g$,min})} ~.
\label{eq:hdinv}
\end{equation}
The left hand side applies to any $d$ dimensional spherically symmetric solution of the equations of motion, in particular to a general non-extremal black hole with horizons at complex radii.
The right hand side is fully determined by the behavior of the potential when the scalars are fixed to attractor values corresponding to the extremal black hole.
The former solution is related to the latter by taking the zero temperature limit.

\section{Invariants for stationary backgrounds}\label{sec:rotation}
We have so far analyzed static spherically symmetric black hole backgrounds in arbitrary two derivative theories of Einstein--Hilbert gravity coupled to matter in arbitrary dimensions.
The analysis was dramatically simplified by the existence of a universal ansatz for these black hole backgrounds which enabled us to focus on a $1+1$ dimensional effective non-linear sigma model and identify the boundary terms that are required to be invariant under a perturbation from one point to another in solution space.
One observes empirically that for known rotating solutions such as the Kerr--Newman solution, there exists a similar invariant constraint on the solution space.
In this case, however, we do not have a universal ansatz for a stationary black hole background, and hence, in order to explore the existence of an invariant in these backgrounds, we adopt a more abstract though generalized analysis.
We sketch the argument below.

In order to restrict to a single angular momentum, we will specialize to the four dimensional case.
Iyer and Wald~\cite{Iyer:1994ys} showed that for any black hole backgrounds with Killing vector $\chi$ in a diffeomorphism invariant theory of gravity, under a perturbation in solution space that keeps the Killing vector invariant, the contribution to the change in the conserved quantity $W=\kappa\, S$ at each Killing horizon is given by 
\begin{equation}\label{6.1}
\delta W = \kappa\, \delta S ~.
\end{equation} 
Here, $\kappa = \chi\cdot \nabla \chi$ evaluated at the Killing horizon, and $S$ is a quantity defined in terms of the local fields evaluated at the horizon and is obtained from variation of the Lagrangian with respect to the Riemann tensor.
At the Schwarzschild horizon, $S$ is the Wald entropy of the black hole background.
The Wald formula can be used to evaluate $S$ at each of the other Killing horizons in the spacetime, except that it no longer carries the interpretation of entropy.

Therefore, given $n$ Killing horizons, we can define the boundary term at each horizon, which depends on the local values of the fields at the $i$-th boundary and the norm of the gradient of the Killing vector at that boundary as
\begin{equation}
W_i = \kappa_i S_i = \alpha_i l_i(\phi) ~,
\end{equation}
where $l_i(\phi)$ is a function of all the fields in the theory evaluated at the Killing horizon. 
Noting that $(a^2)'$ is a temperature in~\eref{6}, this equation is in analogy to~\eref{eq:lpm}. 
Demanding the net variation of the boundary terms under an on-shell variation to be zero, akin to the aforementioned Iyer--Wald perturbation in solution space that gave rise to (\ref{6.1}), we end up with the constraint
\begin{equation}
\sum_i \delta W_i = \sum_i \kappa_i\, \delta S_i = \sum_i \alpha_i l_i(\phi)\, \delta( \log S_i ) = 0 ~. \label{eq:stationary}
\end{equation}
This expression is analogous to what we did in going from~\eref{eq:var2} to~\eref{c0} using~\eref{eq:lpm} or from~\eref{2} to~\eref{2b} using~\eref{3}.
Notice that this is essentially a generalization of the null variation of boundary terms at the Killing horizons for on-shell variations for static black backgrounds, as we have explicated in Section~\ref{sec:cplx}.
It may, however, be useful to step back and examine the inputs that inform~\eref{eq:stationary}.
We have assumed the validity of the first law of black hole mechanics based on the general assumptions of~\cite{Iyer:1994ys}.
We expect that there is a well defined, continuous space of solutions that includes the extremal point and that the variational problem makes sense on this geometry.
To make contact with our discussion of Reissner--Nordstr\"om-like hot attractors, we as well keep any gauge charges fixed when we vary below from one solution to another.
Motivated by the ``no hair'' theorem, we should also keep angular momenta fixed.
Depending on the circumstances under consideration, these assumptions may not be the most general or may be too restrictive.
For our purposes, these are minimal assumptions, and we will nevertheless proceed with these caveats in mind.

Now, in analogy to \eqref{c1}--\eqref{c3}, we consider an on-shell variation between two infinitesimally separated points in solution space, $*$ and $**$. Under a variation from the former to the latter, we obtain the constraint,
\begin{equation}\label{comp1}
 \sum_i \alpha_i l_{i,*}(\phi) \delta(\log \, S_i)=0\,, 
\end{equation}
whereas a reverse variation from $**$ to $*$ yields, 
\begin{equation}\label{comp2}
\sum_i \alpha_i l_{i,**}(\phi) \delta(\log \, S_i)=0\,.
\end{equation}
The two variations \eqref{comp1} and \eqref{comp2} are compatible if and only if 
\begin{equation}
\frac{l_{i,*}(\phi)}{l_{i,**}(\phi)}=\frac{l_{j,*}(\phi)}{l_{j,**}(\phi)} ~,
\end{equation}
with $1\leq i,j\leq n$.
Hence, the $l_i(\phi)$ are constant throughout the space of stationary black hole backgrounds assuming that this solution space is connected. 
The on-shell variational constraint then reduces to 
\begin{equation}\label{ver}
\sum_i \delta ( \log S_i^{l_i(\phi)} ) =0 ~.
\end{equation}
This implies that in theories of gravity in arbitrary dimensions, with smoothly connected solution spaces, determining the $l_i(\phi)$ at any given point fixes them throughout the space. 

In particular, if this space contains a point corresponding to a constant scalar field background that resembles the Kerr--Newman spacetime, then an empirical observation of this solution shows that $l(\phi) = |l_i(\phi)|$ for each Killing horizon $i$.
We therefore arrive at the general result that under an on-shell variation of stationary black hole backgrounds with $n$ Killing horizons in two derivative theories of gravity, in arbitrary dimensions, with smoothly connected solution-spaces, containing a Kerr--Newman background, there exists a conserved quantity given by 
\begin{equation}\label{ver2}
\prod^n_i S_i = f(Q) ~,
\end{equation}
where $f(Q)$ represents a function of all of the non-mass parameters (such as $U(1)$ charges and angular momenta) of the black hole background that are fixed under the on-shell variation. 
We note that the statement (\ref{ver2}), though a stronger statement than (\ref{ver}), can only be made in the context of two derivative theories where empirical constant scalar black hole backgrounds can be explicitly written down, while (\ref{ver})  holds in greater generality.
If a constant scalar Kerr--Newman-like solution exists in a higher derivative theory, we can make the statement~\eref{ver2} in this context as well.

We conclude this section by sketching a heuristic motivation for the area law for rotating, charged black holes in four dimensional ungauged supergravity, and make some contextual comments on the logic of our calculations.
Given an asymptotically flat black hole background with a specific set of charges and angular momenta and with two horizons corresponding to two length scales, if the extremal limit exists, then the non-extremal solution can be regarded as a non-extremal excitation of the extremal background that is characterized by the warp factor $b_\mathrm{ext}(r,\theta)$ and the non-extremality parameter $\Delta$.
This is an uncontroversial observation about the near extremal solutions, but the statement applies more broadly to any non-extremal black hole which admits a smooth zero temperature limit in ${\cal N}=2$ supergravity.
The claim is justified by the attractor mechanism.

We note that~\eref{ver2} tells us that the product of the $S_i$ is an invariant.
Each of the $S_i$ is characterized by a length scale, $b_i$.
In the case where there are two Killing horizons, we have $b_+ (r,\theta)$ and $b_-(r,\theta)$, which, respectively, corresponding to the event and Cauchy horizons.
On dimensional grounds, the two parameterizations are related by 
\begin{equation}
b_+ b_- = \sum_n c_n b^n_\mathrm{ext} \Delta^{2-n} ~.
\end{equation}
Here, we assume that a nice expansion exists for which $c_n$ are system constants; by virtue of extremality, $b_\mathrm{ext}$ is mass independent.
If the double extremal limit exists as a Kerr--Newman solution, we see that $c_n = \delta_{n,2}$.
This motivates the area law for this specific set of black hole backgrounds.

Notice that both the above motivational argument for solutions in the Kerr--Newman class as well as the rigorous discussions for solutions in the Reissner--Nordstr\"om class are motivated by an exploration of the attractor mechanism for black holes. This mechanism, originally discovered for extremal black hole backgrounds with neutral scalar fields, fixes the scalars at the extremal horizon in terms of the charges of the black hole, irrespective of their asymptotic values. 
The on-shell variation in the solution space is therefore naturally performed at fixed charges and with varying values of asymptotic scalar moduli. The entropy of the black hole, which is essentially the horizon length scale is therefore completely independent of all asymptotic data and is purely a function of the charges and the angular momenta. This is the area law for extremal black holes --- a length scale defined purely by non-mass parameters of the black hole. We identified a generalized version of the attractor mechanism under a similar variation for non-extremal charged black holes, and proved a corresponding area law, a product of the horizon length scales which again proved to be defined purely in terms of the non-mass parameters of the black hole. The on-shell variation involves modulating the asymptotic data, which includes the scalar field asymptotia at fixed charges and angular momenta. The ADM mass of the black hole, in the event, that it is defined is purely a function of the scalar asymptotia at fixed charges and angular momenta and its variation is fully specified by the variation in the scalar field data. All the results in this paper are based on this variation, which is natural to the attractor mechanism. Under a different variation in solution space, with say varying charges, one might potentially find different invariants. However, that is an open question at this point in time.

\section{Summary and discussion}\label{sec:disc}


${\cal N}=2$ supergravity supplies a powerful framework for studying black holes.
In particular, in four and five dimensions, the attractor mechanism explains the dynamics of moduli in extremal geometries.
The presence of an AdS$_2$ throat fixes scalars at the horizon.
For non-extremal black holes, our previous work~\cite{gjn1, gjn2} derives conserved quantities associated to the product of the areas of the outer and inner horizon and the extremal attractor equations integrated over the inter-horizon domain.
In this paper, we extend this analysis to arbitrary dimension.
Let us review what we have done.

A change of coordinates~\eref{eq:rho} recasts the equations of motion into a form analogous to what we encounter in four dimensions.
This enables us to establish averaged attractor equations~\eref{hot1} and~\eref{hot2} in higher dimensional settings.
The $d\to\infty$ limit of extremal solutions from Section~\ref{sec:infd} shows that the scalar flow follows a step function that interpolates between the asymptotic moduli and the attractor value at the horizon.
We consider fluctuations about the attractor point in Section~\ref{subsec:pert} and demonstrate the existence of an AdS$_2$ in the space of solutions.
Section~\ref{subsec:exact} and Appendix~\ref{ap:exact} constructs an exact non-extremal solution with a single scalar field that couples to a pair of gauge fields in arbitrary dimension.

In addition to ungauged supergravity, we can introduce Fayet--Iliopoulos terms to perform an Abelian gauging.
At the level of the low energy effective action, the Fayet--Iliopoulos parameters introduce a potential $V_g(\phi)$ for the scalars that acts like a negative cosmological term~\eref{eq:lameff}.
We write the attractor equations in the presence of this term.
Structurally,~\eref{eq:neweom1}--\eref{eq:neweom3} are similar to the equations in the four dimensional case.
The solutions to these equations that we study are the simplest ones:
the scalars are fixed to their attractor values.
In Section~\ref{subsec:vgnonzero}, we demonstrate that there are at most two real strictly positive roots where horizons are situated.

In Section~\ref{sec:cplx}, we examine the horizon invariants more closely.
In ungauged supergravity, following~\cite{gjn1}, it is straightforward to extend the area law to higher dimensions.
With $V_g(\phi)\ne 0$, for constant scalar solutions, the product of the positions of the horizons of a non-extremal black hole is an invariant expressed in terms of the potentials, but the extremal horizon area is not the geometric mean of the areas of the non-extremal horizons.
We establish the form of the invariant~\eref{eq:hdinv} in arbitrary dimension.
This relation applies to the case of $N$ horizons positioned at complex values of the radial coordinate.
The logic used in deriving this expression generalizes to the case of rotating Kerr--Newman-like black holes.
In Section~\ref{sec:rotation}, we sketch this argument in four dimensional ungauged supergravity.

Promoting black hole thermodynamics to black hole statistical physics, we expect that the entropy, which at leading order is computed by the area of the event horizon in Planck units, counts the microstates of the black hole in quantum gravity.
For non-extremal black holes in four dimensions, there are three distinct regions in the spacetime:
Region~1 in which $r\in [r_+,\infty)$, Region~2 in which $r\in [r_-,r_+)$, and Region~3 in which $r\in [0,r_-)$.
In each of the three regions, the time coordinate $t\in (-\infty,\infty)$, but as we cross the horizons sequentially, radial and temporal motion exchange roles in the metric.
We have argued in~\cite{gjn2} that the decoupling of Region~2 from the asymptopia is in analogy to what happens when we take a near horizon limit of the extremal solution.
The AdS$_2$ factor that gets identified in this limit explains how moduli are fixed by the attractor mechanism.
It is telling that the same quantities that vanish by virtue of the attractor equations in the extremal case also vanish when averaged over Region~2 of non-extremal solutions.

The area of the event horizon is determined by $r_+$ and fixes the entropy of the black hole.
The area of the Cauchy horizon is likewise determined by $r_-$.
Assuming the Bekenstein--Hawking formula, the area law instructs us that
\be
S_- = \frac{S_\mathrm{ext}^2}{S_+} ~.
\ee
Since the two are inversely proportional, as we become more and more non-extremal, $S_+$ grows and $S_-$ correspondingly shrinks.
Moreover, as the solution becomes increasingly non-extremal, the reduction in $S_-$ may suggest that there is a universality to states enumerated by the inner horizon.
If the entropy of the Cauchy horizon isolates a particular subset of the states captured by $S_+$, it would be interesting to learn which ones and what selection criteria to apply when restricting to these states.
If Region~2 and Region~3 in the interior of the black hole are decoupled, the additivity of entropy suggests that the degrees of freedom in Region~2 are captured by the difference in the entropies of the horizons: $S_2 = S_+ - S_- = \frac{S_+^2 - S_\mathrm{ext}^2}{S_+}$.
Note that any association of thermodynamic quantities with the Cauchy horizon in supergravity is na\"{\i}vely suspicious as states in the inter-horizon region cannot see the asymptotic vacuum.
We take our cue from the holographic picture of the black hole as follows. 
Since scalar excitations in this region satisfy the Klein--Gordon equation in an AdS$_2$ spacetime, which can be lifted up to a BTZ black hole, in the holographically dual CFT, 
we associate the right movers to the extremal solution and assume without loss of generality that $0 < T_L \le T_R$. 
In this case, from
\be
S_\pm = \frac{\pi r_\pm}{2G_3} = \frac{\pi^2L}{3} (c_R T_R \pm c_L T_L) ~, \qquad
T_\pm^{-1} = \pm \frac{2\pi r_\pm L^2}{r_+^2-r_-^2} = \frac12 \left( \frac{1}{T_R} \pm \frac{1}{T_L} \right) ~,
\ee
we have the equivalent expressions
\be
\label{S2}
S_2 = \frac{\pi\Delta}{G_3} = \frac23 \pi^2L c_L T_L = -\frac23 \pi^2 L c_L\left( \frac{T_+T_-}{T_+-T_-}  \right)  ~,
\ee
where we use the Brown--Henneaux formula to write the central charges $c_L = c_R = \frac{3L}{2G_3}$ in terms of the gravitational constant and the AdS scale.
We may then associate the modes of one set of oscillators --- \textit{viz.}, the ones at a lower temperature --- to the inter-horizon region.
This is consistent with the idea that exciting the second set of oscillators above the ground state is what takes us away from extremality.
While \eqref{S2} is suggestive, we note that macroscopically, Region~2 and Region~3 are not decoupled.  By this we mean that, we cannot find near horizon limits of each horizon which are independently solutions of the equations of motion.
Moreover, in Region~2, an observer is forced to move radially --- \textit{i.e.}, toward the inner horizon --- so it cannot be said that Region~2 and Region~3 are independent.
Because of entanglement, Hilbert spaces do not factorize across horizons.
While we cannot explain these tensions, perhaps they are artifacts of coarse graining microscopic solutions.  
In any case, because the Cauchy horizon suffers a classical instability, how seriously one should take the notion of the inner horizon's entropy is unclear.
Even more mysteriously, one can apply the Bekenstein--Hawking formula to compute the entropy of horizons at the roots of $a(r)$ on the complex $r$-plane and write a first law for these complex horizons.
The interpretation of the complex areas in semiclassical or quantum gravity of course remains elusive.

\comment{
Suppose we consider the Euclidean continuation of the non-extremal black hole geometry.
In setting $\tau_E = it$, we have cut off the interior (Regions~2 and~3) leaving the Euclidean cigar.
We can do something similar in analytically continuing time in Region~3.
The periodicity of the time circles correspond to the inverse temperatures of the black hole horizons.
In computing the partition function from the Euclidean action,
\be
Z[\beta] = \int [Dg][D\Phi]\ e^{-I[g,\Phi]} ~,
\ee
we integrate over metrics $g_{\mu\nu}$ that are smooth with period $\beta$~\cite{Gibbons:1976ue,Hawking:1978jz}.
Knowing that
\be
\beta F = -\log Z ~, \qquad \langle E \rangle = -\frac{\partial}{\partial\beta} \log Z ~,
\ee
we then compute entropy as a Legendre transform:
\be
S = \beta (\langle E \rangle - F) \approx \left( \beta \frac{\partial}{\partial\beta} - 1 \right) I_\mathrm{cl} ~.
\ee
We have seen that there are invariants associated to non-extremal geometries, for example, $A_\mathrm{ext}^2$, or its generalization in higher dimensions~\eref{eq:hdinv}.
The fact that the complex horizons contribute to black hole invariants suggests that they also make a contribution to the partition function calculated on the Euclidean section.
Future work interrogates how this occurs.
}

In addition to \eqref{eq:blahblah1}, one can use Vieta's formula to extract other interesting horizon relationships from \eqref{eq:ab_sol} for AdS Reissner--Nordstr\"om black holes.\footnote{
Similar relations were derived in~\cite{Wang:2013smb,Xu:2013zpa,Xu:2014qaa,Du:2014kpa}.}
For instance, one finds,
\begin{eqnarray}
\label{Vieta}
&& \sum_i r_i = 0 ~, \qquad
\sum_{i<j} r_i r_j = \frac{(d-1)(d-2)}{2(-V_{g,\mathrm{min}})} ~, \qquad
\sum_{i<j<k} r_i r_j r_k = 0 ~, \qquad
\ldots ~, \cr
&& \sum_{1\le i_1 < i_2 < \ldots < i_{d-1} \le N} r_{i_1}\ldots r_{i_{d-1}} = (-1)^d \frac{8\pi G_d M (d-1)}{\omega_{d-2}(-V_{g,\mathrm{min}})} ~, \qquad \ldots ~.
\end{eqnarray}
We leave an investigation of how these relations generalize in the presence of non-constant scalars to future work.

\section*{Acknowledgements} 
We thank Paolo Benincasa, Gabriel Cardoso, Alejandra Castro,  Mirjam Cvetic, Roberto Emparan, Finn Larsen, Shiraz Minwalla, Shahin Sheikh-Jabbari, Kostas Skenderis, Marika Taylor, and Sandip Trivedi for comments and discussions.
KG and VJ are supported in part by the South African National Research Foundation. 
VJ and JM are supported by the South African Research Chairs Initiative of the Department of Science and Technology.
JM is also supported by the National Institute for Theoretical Physics.
SN is supported by the FCT fellowship FCT-DFRH-Bolsa SFRH/BPD/101955/2014.
The authors are grateful to each other's home institutions for supporting reciprocal visits.

\appendix

\section{Exact solutions}
\label{ap:exact}

In this Appendix, we go through the calculations used to find the solution \eqref{eq:simpler2}.
To this end, it is convenient to use the variables
\be
u_1 = \phi ~, \qquad u_2 = \log a ~, \qquad v = \log a b^{d-3} ~.
\ee
We define
\be
\dot{} = \partial_\tau = e^{2v} \partial_\rho
\ee
so that
\be
\tau = \int \frac{d\rho}{a^2 b^{2(d-3)}} = \frac{1}{\rho_+-\rho_-} \log\frac{\rho-\rho_+}{\rho-\rho_-} ~.
\label{eq:TAU}
\ee
Equivalently,
\be
\rho = \frac{\rho_+ - e^{(\rho_+-\rho_-)\tau} \rho_-}{1-e^{(\rho_+-\rho_-)\tau}} ~. \label{eq:rhotau}
\ee

In terms of the new variables, we recast~\eref{dilatoneqhd},~\eref{ham_hd},~\eref{eq:hd_1}, and~\eref{eq:bdashdash} as, respectively,
\bea
\ddot{u}_1 &=& \frac{(d-2)!}{4(d-3)^2} \sum_a \alpha_a e^{2u_2+\alpha_a u_1} (Q_m^a)^2 ~, \label{eq:u1} \\
\ddot{u}_2 &=& (d-4)! \sum_a e^{2u_2+\alpha_a u_1} (Q_m^a)^2 ~, \label{eq:u2} \\
\ddot{v} &=& e^{2v} ~, \label{eq:veq} \\
\dot{v}^2 - \ddot{v} &=& \frac{2(d-3)}{d-2} \dot{u}_1^2 + \dot{u}_2^2 - (d-4)!\sum_a e^{2u_2+\alpha_a u_1} (Q_m^a)^2 ~. \label{eq:newencont}
\eea
From~\eref{eq:ab_hd}, we have
\be
e^{2v} = (\rho-\rho_+)(\rho-\rho_-) ~. \label{eq:vee}
\ee
Using~\eref{eq:rhotau}, the expression for $v$ in~\eref{eq:vee} tells us that the left hand side of~\eref{eq:newencont} is independent of $\tau$:
\be
\dot{v}^2 - \ddot{v} = \tfrac{1}{4}(\rho_+-\rho_-)^2 =: {\cal E} ~.
\ee
Now, defining
\be
X_a = \sum_b \big( n_{ab}^{-1} u_b + m_{ab}^{-1} \log[(\alpha_1-\alpha_2)(Q_m^b)^2] \big) ~,
\ee
where
\be
n^{-1} = \left( \ba{cc} \frac{4(d-3)^2}{(d-2)!} & -\frac{\alpha_2}{(d-4)!} \cr -\frac{4(d-3)^2}{(d-2)!} & \frac{\alpha_1}{(d-4)!} \ea \right) ~, \qquad
m_{ab} = \frac{(d-4)!}{4(\alpha_1-\alpha_2)}\left(8+\frac{d-2}{d-3}\alpha_a \alpha_b\right) ~,
\ee
we find that~\eref{eq:u1} and~\eref{eq:u2} may be rewritten as
\be
\ddot{X}_a = e^{m_{ab} X_b} ~, \label{eq:ddotx}
\ee
subject to the constraint~\eref{eq:newencont}
\be
\frac12 \dot{X}_a m_{ab} \dot{X}_b - \sum_a e^{m_{ab} X_b} = \frac{\alpha_1-\alpha_2}{(d-4)!}{\cal E} ~.
\ee
The $\tau$-evolution of $v$ given in~\eref{eq:veq} decouples from the equations of motion for the $X_a$.

In order to decouple the equations~\eref{eq:ddotx} for $X_a$, we take
\be
8 + \frac{d-2}{d-3} \alpha_1 \alpha_2 = 0 ~.
\ee
Without loss of generality, assume that $\alpha_1 > 0 > \alpha_2$.\footnote{
	Plugging in to~\eref{eq:gammai}, this is the case $\gamma=1$.}
This implies that
\be
m_{ab} = \frac{(d-2)!}{4(d-3)^2} \mathrm{diag}(\alpha_1,-\alpha_2) ~.
\ee
The differential equation~\eref{eq:ddotx} has a general solution
\be
X_a = \frac{1}{|\alpha_a|\kappa} \log\frac{2c_a^2}{|\alpha_a|\kappa \sinh^2(c_a(\tau-d_a))} ~, \qquad
\kappa := \frac{(d-2)!}{4(d-3)^2} ~,
\ee
where the $c_a$ and $d_a$ are integration constants.

\comment{
	Noting that
	\be
	\frac{\partial}{\partial\phi} V_\mathrm{eff}(\phi) = 0 \qquad \Longrightarrow \qquad \phi_0 = -\frac{\log\left(-\frac{\alpha_1 (Q_m^1)^2}{\alpha_2 (Q_m^2)^2}\right)}{\alpha_1-\alpha_2} ~,
	\ee
	we have, from~\eref{eq:gammai},
	\be
	\gamma = \frac12 \left( \left[ 1+ \frac{d-2}{d-3} (-\alpha_1 \alpha_2) \right]^\frac12 - 1 \right) = 1 ~.
	\ee
}

Now, the scalar field and the warp factors that appear in the metric are written in terms of the charges and the coordinate $\tau(\rho)$ as
\bea
e^{(\alpha_1-\alpha_2)\phi} &=& \frac{(Q_m^2)^2}{(Q_m^1)^2}\; e^{\kappa \sum_a \alpha_a X_a} ~, \label{eq:sceq} \\
a^2 &=& \exp\left( \frac{2(d-4)!}{\alpha_1-\alpha_2} (X_1+X_2) \right) / \Diamond ~, \\
b^{2(d-3)} &=& \frac{(\rho-\rho_+)(\rho-\rho_-)}{a^2} ~,
\eea
where
\be
\Diamond = (\alpha_1-\alpha_2) (Q_m^1)^{\frac{-2\alpha_2}{\alpha_1-\alpha_2}} (Q_m^2)^{\frac{2\alpha_1}{\alpha_1-\alpha_2}} ~.
\ee
Thus, in particular,~\eref{eq:sceq} tells us that
\be
\label{phi_gen}
e^{(\alpha_1-\alpha_2)\phi} = -\frac{\alpha_2 c_1^2 (Q_m^2)^2}{\alpha_1 c_2^2 (Q_m^1)^2}\; \frac{\sinh^2(c_2(\tau-d_2))}{\sinh^2(c_1(\tau-d_1))} ~.
\ee
Because $\tau\to-\infty$ as $\rho\to \rho_+$, finiteness of the solution at the horizon implies $c_1=c_2=:c$.
Using~\eref{eq:TAU} and the scaling in the vicinity of the outer horizon, we also see that
\be
b^{2(d-3)} \sim \frac{\rho-\rho_+}{a^2} \sim e^{(\rho_+-\rho_-)\tau-2c\tau} ~.
\ee
We therefore discover that $c = \frac12(\rho_+-\rho_-)$ in order for the horizon area to be finite.
This is $\Delta$, the non-extremality parameter. 
It serves to define our solution.
Similarly, the flat space asymptotic boundary conditions tell us that
\bea
&& \Diamond = \exp\left[ -\frac{\alpha_2}{\alpha_1-\alpha_2}\log\left(\frac{-\alpha_2 c^2}{(d-4)!\sinh^2(cd_1)}\right) + \frac{\alpha_1}{\alpha_1-\alpha_2}\log\left(\frac{\alpha_1 c^2}{(d-4)!\sinh^2(cd_2)}\right) \right] ~, \nn \\ && \\
&& e^{(\alpha_1-\alpha_2)\phi_\infty} = -\frac{\alpha_2(Q_m^2)^2}{\alpha_1(Q_m^1)^2}\; \frac{\sinh^2(cd_2)}{\sinh^2(cd_1)} ~.
\eea
The first equality arises from the condition $a^2\to 1$ as $\rho\to\infty$ (or $\tau\to 0$).
We may thus solve for the unknown parameters $d_a$ in terms of the charges and the asymptotic value of the scalar, $\phi_\infty$.

For definiteness, let us restrict to the special case $\alpha_1=-\alpha_2=\alpha = \sqrt\frac{8(d-3)}{d-2}$.
For this solution,
\be
\label{eq:sinh_ids}
\sinh c d_1 =  \frac{c}{\sqrt{2(d-4)!}\, \bar Q_m^1} ~, \qquad
\sinh c d_2 =  \frac{c}{\sqrt{2(d-4)!}\, \bar Q_m^2} ~,
\ee
where we recall that $\bar Q_m^a$ was defined in \eqref{Qdef}.
Thus,
\be
e^{\alpha\phi} = e^{\alpha\phi_\infty} 
\frac{\bar Q_m^2 \sinh\left(\log e^{-cd_2}\big(\frac{\rho-\rho_+}{\rho-\rho_-}\big)^\frac{1}{2}\right)}{\bar Q_m^1 \sinh\left(\log e^{-cd_1}\big(\frac{\rho-\rho_+}{\rho-\rho_-}\big)^\frac{1}{2}\right)} ~.
\label{eq:thus}
\ee
Using the identity $\sinh \log X=\frac{X^2-1}{2X}$ and~\eqref{eq:sinh_ids}, we obtain 
\begin{equation}
\label{eq:simpler}
e^{\alpha\phi} = e^{\alpha\phi_\infty} 
\frac{\rho-f_2}{\rho-f_1} ~,
\end{equation}
where
\begin{equation}
\label{eq:f_def}
f_a =  \frac{e^{cd_a}\rho_- - e^{-cd_a}\rho_+}{2\sinh (c d_a) } ~.
\end{equation}
Also,
\begin{eqnarray}
\label{eq_br}
b^{2(d-3)}&=&a^{-2}(\rho-\rho_+)(\rho-\rho_-) \cr
&=& \tfrac{2\bar Q_m^2 \bar Q_m^1(d-4)!}{c^2} 
\sinh\left(\log e^{-cd_2}\left(\tfrac{\rho-\rho_+}{\rho-\rho_-}\right)^\frac{1}{2}\right)
\sinh\left(\log e^{-cd_1}\big(\tfrac{\rho-\rho_+}{\rho-\rho_-}\big)^\frac{1}{2}\right)(\rho-\rho_+)(\rho-\rho_-)\cr
&=& (\rho - f_1 )(\rho- f_2) ~.
\end{eqnarray}

Assuming that we are working in an asymptotically flat space, $b\rightarrow r$ at spatial infinity, so from~\eqref{eq:rho}, $\rho\rightarrow r^{d-3}$ and from~\eqref{eq:ab_hd}
\begin{equation}
\label{M}
a^2 \rightarrow \frac{(r^{d-3}-\rho_+)(r^{d-3}-\rho_-)}{r^{2(d-3)}} \approx 1-\frac{\rho_++\rho_-}{r^{d-3}}+\ldots ~.
\end{equation}
We are now able to read off the mass parameter from the metric obtaining \eqref{Mrho}
Asymptotically, we also have
\begin{equation}
\label{eq:tau_far}
\partial_\tau \rightarrow \rho^2\partial_\rho ~.
\end{equation}
Examining~\eref{M}, this yields
\begin{equation}
\label{label}
M=\dot a |_{\tau=0} ~,
\end{equation}
and likewise,
\begin{equation}
\label{eq:l_def}
\Sigma = -\dot\phi|_{\tau=0} ~.
\end{equation}
Using these definitions,~\eqref{eq:newencont} tells us that
\begin{equation}
\label{label}
\tfrac{1}{4}\alpha^2 \Sigma^2 +M^2 -(d-4)!\left((\bar{Q}^1_m)^2 +  (\bar{Q}^2_m)^2\right)  = c^2 ~.
\end{equation}
Taking the $\tau$ derivative of~\eref{eq:thus} and the expression for $a^2$ from~\eref{eq_br}, we find
\begin{eqnarray}
\label{sigma}
\tfrac{1}{2}\alpha\Sigma &=& \tfrac{1}{2} c ({\coth(cd_2)-\coth(cd_1)}) ~, \\
\label{m2}
M &=& \tfrac{1}{2}c (\coth(cd_2)+\coth(cd_1)) ~.
\end{eqnarray}
Combining~\eqref{sigma} and~\eqref{m2}, we then have
\begin{equation}
\label{ms}
\alpha M \Sigma = \tfrac{1}{2} c^2 \left(\text{csch}^2(cd_2)-\text{csch}^2(cd_1)\right) = (d-4)!((\bar{Q}_m^2)^2-(\bar{Q}_m^1)^2) ~.
\end{equation}
Since
\begin{equation}
\label{eq:f_def2}
f_1 = \tfrac{1}{2}\left(\rho_+ + \rho_- - (\rho_+-\rho_-)\coth(c d_1)\right) = M - c \coth(c d_1) = \tfrac{1}{2}\alpha \Sigma ~,
\end{equation}
and similarly $f_2=-\frac{1}{2}\alpha\Sigma$, we get \eqref{eq:simpler2}

Substituting the expressions for $f_a$ in~\eref{eq_br}, we recast the warp factor in the metric as \eqref{b_simpler}
\subsection{Extremal limit}
To find the extermal solution, one takes the limit $c\rightarrow0$, and from \eqref{phi_gen}, with $\alpha=\alpha_1=-\alpha_2$, one finds
\begin{equation}
\label{extremal_D}
e^{\alpha\phi} = \frac{|Q_m^2|}{ |Q_m^1|}\; \frac{\tau-d_2}{\tau-d_1} ~.
\end{equation}
Taking the  $c\rightarrow 0$ limit of \eqref{eq:sinh_ids} 
one gets
\be
\label{eq:sinh_ids2}
d_1 =  \frac{1}{\sqrt{2(d-4)!}\, \bar Q_m^1} ~, \qquad
d_2 =  \frac{1}{\sqrt{2(d-4)!}\, \bar Q_m^2} ~,
\ee
so that 
\begin{equation}
\label{extremal_D2}
e^{\alpha\phi} = e^{\alpha\phi_\infty}\frac{\sqrt{2(d-4)!}|\bar Q_m^2|\tau-1}{\sqrt{2(d-4)!} |\bar Q_m^1|\tau-1}~.
\end{equation}
Note, that with $c=0$, $\rho_+=\rho_-=M$, so from \eqref{eq:TAU} we obtain
\begin{equation}
\label{tau_ext}
\tau=-\frac{1}{\rho-M} 	 
\end{equation}
\begin{equation}
\label{extremal_D3}
e^{\alpha\phi} = e^{\alpha\phi_\infty}\frac{\sqrt{2(d-4)!}|\bar Q_m^2|+(\rho-M)}{\sqrt{2(d-4)!} |\bar Q_m^1|+(\rho-M)}~.
\end{equation}
Taking the $c\rightarrow0$ limit of \eqref{sigma} and \eqref{m2}
\begin{eqnarray}
\label{sigma_ex}
\tfrac{1}{2}\alpha\Sigma &=& \tfrac{1}{2}  ({(d_2)^{-1}-(d_1)^{-1}})=\sqrt{\frac{(d-4)!}{2}}\left(\bar Q_m^2-\bar Q_m^1\right) ~, \\
\label{m_x}
M &=& \tfrac{1}{2} ((d_2)^{-1}+(d_1)^{-1}) =\sqrt{\frac{(d-4)!}{2}}\left(\bar Q_m^2+\bar Q_m^1\right) ~.
\end{eqnarray}
and we see that in fact \eqref{extremal_D3} is the same as \eqref{eq:simpler2}.
As a check, we note that taking the  $c\rightarrow 0$ limit of \eqref{eq_br} gives
\begin{eqnarray}
\label{b_ext}
b^{2(d-3)}&=&2 Q_m^1 Q_m^2 (d-4)!(\tau-d_1)(\tau-d_2)(\rho-M)^2 \\
&=& (\rho-M-\sqrt{2(d-4)!}|\bar Q_m^2|)(\rho-M-\sqrt{2(d-4)!}|\bar Q_m^1|)\\
&=& \rho^2-(\tfrac{1}{2}\alpha\Sigma)^2
\end{eqnarray}

\bibliographystyle{JHEP}
\bibliography{bibfile}

\end{document}